%% file: B2G-16-004_temp.tex
\pdfoutput=1

\documentclass[11pt,twoside,a4paper,cmspaper,final,collab]{cms-tdr}
\def\svnVersion{398794}\def\cmsCernNoTag{CERN-EP/2016-296}\def\cmsCernDate{\today}\def\cmsMessage{Published in the Journal of High Energy Physics as \href{http://dx.doi.org/10.1007/JHEP03(2017)162}{\doi{10.1007/JHEP03(2017)162.}}}

\begin{document}\cmsNoteHeader{B2G-16-004}

\hyphenation{had-ron-i-za-tion}
\hyphenation{cal-or-i-me-ter}
\hyphenation{de-vices}
\RCS$Revision: 398649 $
\RCS$HeadURL: svn+ssh://svn.cern.ch/reps/tdr2/papers/B2G-16-004/trunk/B2G-16-004.tex $
\RCS$Id: B2G-16-004.tex 398649 2017-04-11 11:25:08Z jngadiub $
\newlength\cmsFigWidth
\setlength\cmsFigWidth{0.48\textwidth}
\cmsNoteHeader{B2G-16-004}

\newcommand{\PV}{\ensuremath{\mathrm{V}}}
\providecommand{\NA}{---}
\newcommand{\LOWSBLOW}{40\xspace}
\newcommand{\SRLOW}{65\xspace}
\newcommand{\SRMIDDLE}{85\xspace}
\newcommand{\SRHIGH}{105\xspace}
\newcommand{\HIGHSBLOW}{135\xspace}
\newcommand{\HIGHSBHIGH}{150\xspace}
\newcommand{\SFWTAGHPWPT} {\ensuremath{0.95\pm0.06}\xspace}
\newcommand{\SFWTAGLPWPT} {\ensuremath{1.25\pm0.32}\xspace}
\newcommand{\SFWTAGHPWPL} {\ensuremath{1.01\pm0.03}\xspace}
\newcommand{\WMASSDATAWPT} {\ensuremath{84.6\pm0.7}\xspace}
\newcommand{\WRESDATAWPT} {\ensuremath{8.2\pm0.7}\xspace}
\newcommand{\WMASSMCWPT} {\ensuremath{85.1\pm0.2}\xspace}
\newcommand{\WRESMCWPT} {\ensuremath{7.8\pm0.3}\xspace}
\newcommand{\LUMIUNCERT} {2.7\%\xspace}
\newcommand{\Wo}{\PW\xspace}%
\newcommand{\Zo}{\cPZ\xspace}%
\newcommand{\Vo}{\ensuremath{\mathrm{V}}\xspace}%
\newcommand{\mX}{\ensuremath{\text{M}_{\text{X}}}}%
\newcommand{\wX}{\ensuremath{\Gamma_{\text{X}}}}%
\newcommand{\mVV}{\ensuremath{m_{\Vo\Vo}}\xspace}%
\newcommand{\mjj}{\ensuremath{m_\mathrm{jj}}}%
\newcommand{\mJ}{\ensuremath{m_{\text{jet}}}}%
\newcommand{\nsubj}{\ensuremath{\tau_{21}}}%
\newcommand{\ktilde}{\ensuremath{k/\overline{M}_\mathrm{Pl}}\xspace}%
\newcommand{\amcatnlo}{{\MADGRAPH{}5\_a\MCATNLO}\xspace}
\newcommand{\BulkG}{\ensuremath{\PXXG_{\text{bulk}}}\xspace}
\newcommand{\lnujet}{\ensuremath{\ell \nu}+jet\xspace}
\newcommand{\Wpr}{\ensuremath{\rm W^\prime}\xspace}
\newcommand{\Zpr}{\ensuremath{\rm Z^\prime}\xspace}
\renewcommand{\PTm}{\ensuremath{p_{\mathrm{T}}^{\text{miss}}}\xspace}
\newcommand\T{\rule[-.50em]{0pt}{1.50em}}

\title{Search for massive resonances decaying into WW, WZ or ZZ bosons in proton-proton collisions at $\sqrt{s} = 13$\TeV}

\author[unesp]	{	Sudha	Ahuja	}
\author[ttu]    {   Nural	Akchurin	}
\author[uzh]	{	Thea Klaeboe Aarrestad	}
\author[milano]	{	Luca	Brianza	}
\author[ncu]	{	Yu-Hsiang Chang	}
\author[ncu]	{	Ching-Wei Chen	}
\author[ttu]	{	Jordan	Damgov	}
\author[ttu]	{	Phil	Dudero	}
\author[ulb]    {       Laurent Favart  }
\author[milano]	{	Raffaele Gerosa	}
\author[milano]	{	Alessio Ghezzi	}
\author[lyon]	{	Maxime	Gouzevitch	}
\author[milano]	{	Pietro  Govoni	}
\author[fnal]	{	Lindsey Gray 	}
\author[uzh]	{	Andreas	Hinzmann	}
\author[pku]    {	Huang   Huang	}
\author[ncu]	{	Ji-Kong	Huang	}
\author[ncu]	{	Raman	Khurana	}
\author[uzh]	{	Ben	Kilminster	}
\author[uzh]	{	Clemens	Lange	}
\author[ttu]	{	Sung-Won	Lee	}
\author[pku]	{	Qiang	Li	}
\author[ncu]	{	Yun-Ju	Lu	}
\author[jhu]	{	Petar	Maksimovic	}
\author[madrid]	{	Dermot  Moran	}
\author[kit]	{	Matthias Mozer	}
\author[uzh]	{	Jennifer	Ngadiuba	}
\author[unesp]	{	Sergio	Novaes	}
\author[padova]	{	Alexandra	Oliveira	}
\author[padova]	{	Jacopo Pazzini	}
\author[caltech] { Maurizio Pierini}
\author[buffalo]	{	Salvatore	Rappoccio	}
\author[unesp]  {       Jos\'e Ruiz     }
\author[kit]    {       Daniela Sch\"afer       }
\author[unesp]	{	Thiago	Tomei	}
\author[ncu]	{	Henry Yee-Shian	Tong	}
\author[fnal]	{	Nhan	Tran	}
\author[madrid]	{	Jorge   Troconiz	}
\author[pku]	{	Qun	Wang	}
\author[pku]	{	Mengmeng	Wang	}
\author[ncu]	{	Jun-Yi	Wu	}
\author[pku]	{	Zijun	Xu	}
\author[ncu]	{	Shin-Shan Eiko	Yu	}
\author[pku]	{	Xiao-Qing Yuan	}
\author[padova]	{	Alberto Zucchetta	}

\date{\today}

\abstract{
A search is presented for new massive resonances decaying to WW, WZ or ZZ bosons in $\ell\Pgn\qqbar$ and $\qqbar\qqbar$ final states.
Results are based on data corresponding to an integrated luminosity of 2.3--2.7\fbinv recorded in proton-proton collisions at $\sqrt{s} = 13$\TeV with the CMS detector at the LHC. Decays of spin-1 and spin-2 resonances into two vector bosons are sought in the mass range 0.6--4.0\TeV.
No significant excess over the standard model background is observed.
Combining the results of the $\ell\Pgn\qqbar$ and $\qqbar\qqbar$ final states, cross section and mass exclusion limits
are set for models that predict heavy spin-1 and spin-2 resonances.
This is the first search for a narrow-width spin-2 resonance at $\sqrt{s} = 13$\TeV.
}

\hypersetup{%
pdfauthor={CMS Collaboration},%
pdftitle={Search for massive resonances decaying into WW, WZ or ZZ bosons in proton-proton collisions at sqrt(s) = 13 TeV},%
pdfsubject={CMS},%
pdfkeywords={CMS, B2G, resonances}}

\maketitle
\section{Introduction}
\label{sec:introduction}

Several theories beyond the standard model (SM) predict the existence of heavy particles that preferentially decay to pairs of vector bosons \PV{}, where \PV{} represents a W or Z. These models usually aim to clarify open questions in the SM such as the apparently large difference between the electroweak and the gravitational scales.
Notable examples of such models include the bulk scenario~\cite{Agashe:2007zd, Fitzpatrick:2007qr, Antipin:2007pi} of the Randall--Sundrum warped extra-dimensions (RS1)~\cite{Randall:1999ee,Randall:1999vf} and a heavy vector-triplet (HVT) model~\cite{Pappadopulo:2014qza}.
The bulk graviton model is described by two free parameters: the mass of the first Kaluza-Klein (KK) excitation of a spin-2 boson (the KK bulk graviton $\mathrm{G}_\text{bulk}$) and
the ratio $\tilde{k} \equiv k/\overline{M}_\mathrm{Pl}$,
where $k$ is the unknown curvature scale of the extra dimension and $\overline{M}_\mathrm{Pl} \equiv M_{Pl}/\sqrt{8\pi}$ is the reduced Planck mass.
The HVT generalises a large number of models that predict spin-1 charged (\Wpr{}) and neutral (\Zpr{}) resonances.
Such models can be described in terms of just a few parameters:
two coefficients $c_\mathrm{F}$ and $c_\mathrm{H}$, scaling the couplings to fermions, and to the Higgs and longitudinally polarized SM vector bosons respectively, and the strength $g_{\PV}$ of the new vector boson interaction.
Two benchmark models are considered in the HVT scenario. In the first one, referred to as HVT model A, with $g_{\PV} = 1$,
weakly coupled vector resonances arise from extensions of the SM gauge group, such as the
sequential standard model (SSM)~\cite{Altarelli}, that have comparable branching fractions to fermions and gauge bosons.
In HVT Model B with $g_{\PV} = 3$, the new resonances have large branching fractions to pairs of bosons, while their fermionic couplings are suppressed.
This scenario is most representative of composite models of Higgs bosons.

Searches for diboson resonances have been previously performed in many different final states, placing lower limits on the masses of these resonances above the TeV scale~\cite{Aaboud:2016lwx,Aaboud:2016okv,Khachatryan:2014xja,Khachatryan:2014gha,Khachatryan:2014hpa,Khachatryan:2016yji,Khachatryan:2015bma,Khachatryan:2015ywa,Aad:2015owa,Aad:2015ufa,ATLASwprimeWZPAS,Aad:2015yza}.
Searches performed with proton-proton collisions at $\sqrt{s} = 8\TeV$ indicated deviations from background expectations at resonance masses of about 2\TeV.
The largest excesses of events were observed in the searches
in the dijet WW, WZ or ZZ~\cite{Aad:2015owa,Khachatryan:2014hpa} channels,
as well as in the semi-leptonic WH $\rightarrow\ell\Pgn\bbbar$ final state~\cite{Khachatryan:2016yji}, and have local significances of 3.4$\sigma$ and 2.2$\sigma$, respectively.
The most stringent lower mass limit for a \Wpr (\Zpr) is set at 2.3 (2.0) \TeV by a combination of searches in semi-leptonic and all-hadronic final states performed with proton-proton collisions at $\sqrt{s} = 13\TeV$~\cite{Aaboud:2016okv}.
The same searches provide the current lower mass limit of 2.6\TeV for a HVT.

This paper presents a search for resonances with masses above 0.6\TeV decaying into a pair of vector bosons.
The analysis is based on data collected in proton-proton collisions at $\sqrt{s} = 13\TeV$ with the CMS experiment at the CERN LHC during 2015, corresponding to an integrated luminosity of 2.3 and 2.7\fbinv for the $\ell\Pgn\qqbar$, where $\ell=\Pgm$ or e, and $\qqbar\qqbar$ final state, respectively. The \lnujet{} search also includes the $\PW\to\Pgt\Pgn$ contribution from the decay $\Pgt\to\ell\Pgn\Pagn$. The gain in sensitivity from $\Pgt$ leptons is limited by the small branching fractions involved.

The key challenge of the analyses is the reconstruction of the highly energetic decay products. Since the resonances under study have masses of order 1\TeV, their decay products, \ie{} the bosons, have on average transverse momenta (\pt) of several hundred GeV or more. As a consequence, the particles emerging from the boson decays are very collimated. In particular, the jet-decay products of the bosons cannot be resolved using the standard algorithms, but are instead reconstructed as a single jet object. Dedicated techniques, called jet ``V tagging'' techniques, are applied to exploit the substructure of such objects, to help resolve jet decays of massive bosons~\cite{CMS-PAS-JME-14-002,Khachatryan:2014vla}. The V tagging also helps suppress SM backgrounds, which mainly originate from the production of multijet, W+jets, and nonresonant VV events.

The final states considered are either $\ell\Pgn\qqbar$ or $\qqbar\qqbar$, where the hadronic decay products of the V decay are
reconstructed in a single jet. They result in events with either a charged lepton, a neutrino, and a single reconstructed jet (\lnujet{} channel), or two reconstructed jets (dijet channel).
As in the analyses of previous data~\cite{Khachatryan:2014gha, Khachatryan:2014hpa}, the aim is to reconstruct all decay products of the new resonance to be able to search for a localized enhancement in the diboson invariant mass spectrum. While the analyses in general aim at large resonance masses, we conduct two exclusive searches in the \lnujet{} final state, separately optimized for the mass ranges 0.6--1.0\TeV (``low-mass'') and $> 1$\TeV (``high-mass'').

This paper is organized as follows. Section~\ref{sec:cmsdetector} briefly describes the CMS detector. Section~\ref{sec:simulatedsamples} gives an overview of the simulations used in this analysis. Section~\ref{sec:eventreconstruction} provides a detailed description of the reconstruction and event selection. Section~\ref{sec:backgroundestimation} describes the background estimation and the signal modelling procedures. Systematic uncertainties are discussed in Section~\ref{sec:systematicuncertainties}. The results of the search for a spin-2 bulk graviton and for spin-1 resonances as predicted by HVT models are presented in Section~\ref{sec:statisticalinterpretation}, and Section~\ref{sec:summary} provides a brief summary.
\section{The CMS detector}
\label{sec:cmsdetector}

The central feature of the CMS apparatus is a superconducting solenoid of 6\unit{m} internal diameter, providing a magnetic field of 3.8\unit{T}. Contained within the solenoid volume are a silicon pixel and strip tracker, a lead tungstate crystal electromagnetic calorimeter (ECAL), and a brass and scintillator hadron calorimeter (HCAL), each composed of a barrel and two endcap sections. Extensive forward calorimetry complements the coverage provided by the barrel and endcap detectors. The forward hadron (HF) calorimeter uses steel as an absorber and quartz fibers as the sensitive material. The two halves of the HF are located 11.2\unit{m} from the interaction region, one on each end, and together they provide coverage in the pseudorapidity range $3.0 < \abs{\eta} < 5.2$.
Muons are measured in gas-ionization detectors embedded in the steel flux-return yoke outside the solenoid.

A particle-flow (PF) event algorithm~\cite{CMS-PAS-PFT-09-001,CMS-PAS-PFT-10-001} reconstructs and identifies each individual particle with an optimized combination of information from the various elements of the CMS detector. The energy of photons is obtained from the ECAL measurement, corrected for suppression of small readout signals. The energy of electrons is determined from a combination of the electron momentum at the primary interaction vertex as determined by the tracker, the energy of the corresponding ECAL cluster, and the energy sum of all bremsstrahlung photons spatially compatible with originating from the electron track. The momentum of muons is obtained from the curvature of the corresponding track. The energy of charged hadrons is determined from a combination of their momentum measured in the tracker and the matching of energies deposited in ECAL and HCAL, also corrected for suppression of small signals and for the response function of the calorimeters to hadronic showers. Finally, the energy of neutral hadrons is obtained from the corresponding corrected ECAL and HCAL energy.

A more detailed description of the CMS detector, together with a definition of the coordinate system and the kinematic variables, can be found in Ref.~\cite{Chatrchyan:2008aa}.
\section{Simulated samples}
\label{sec:simulatedsamples}

The bulk graviton model
and HVT models are used as benchmark signal processes.
In these models, the vector gauge bosons are produced with a longitudinal polarization in more than 99\% of the cases.
For each resonance hypothesis, we consider masses in the range 0.6 to 4.0\TeV.
Simulated signal events are generated at leading order (LO) accuracy with \amcatnlo{} v2.2.2~\cite{Alwall:2014hca} with a width of 0.1\% of the resonance mass.

The Monte Carlo (MC) generated samples of SM backgrounds are used to optimize the analyses.
The W+jets SM process is simulated with \amcatnlo{}, while \ttbar and single top quark events are generated with both \POWHEG v2~\cite{Nason:2004rx,Frixione:2007vw,Alioli:2010xd,Alioli:2009je,Re:2010bp,Alioli:2011as} and \amcatnlo.
Diboson (WW, WZ, and ZZ) processes are generated with \PYTHIA v8.205 \cite{Sjostrand:2006za,Sjostrand:2007gs}.
Parton showering and hadronization are implemented through \PYTHIA using the CUETP8M1 tune~\cite{Skands:2014pea,Khachatryan:2015pea}.
The NNPDF 3.0~\cite{Ball:2011mu} parton distribution functions (PDFs) are used for all simulated samples, except for diboson ones (WW, WZ and ZZ) for which NNPDF 2.3LO is used.
All events are processed through a \GEANTfour-based~\cite{Agostinelli:2002hh} simulation of the CMS detector.
The simulated background is normalized using inclusive cross sections calculated at next-to-leading order (NLO), or next-to-NLO order in quantum chromodynamics (QCD) where available, using \MCFM v6.6 \cite{MCFM:VJets,MCFM:VV,Campbell:2012uf,MCFM:SingleTop} and \FEWZ v3.1 \cite{FEWZ3}.

Additional simulated minimum-bias interactions are added to the generated events to match the additional particle production observed in the large number of overlapping proton-proton interactions within the same or nearby bunch crossings (pileup). The simulated events are corrected for differences between data and simulation in the efficiencies of the lepton trigger~\cite{CMS:FirstInclZ}, lepton identification and isolation~\cite{CMS:FirstInclZ}, and selection of jets originating from hadronization of b quarks (b jets)~\cite{CMS-PAS-BTV-15-001}.
\section{Reconstruction and selection of events}
\label{sec:eventreconstruction}

\subsection{Trigger and preliminary offline selection}

In the \lnujet{} channel, events are collected with a trigger requiring either one muon or one electron.
For the low-mass \lnujet{} analysis, both triggers have a \pt requirement of 27\GeV. The muons and electrons selected online also satisfy both isolation requirements and identification criteria.
The selection efficiency of these triggers for leptons satisfying the offline requirements described in Section~\ref{subsec:leptons}, varies in the range $90$--$95$\% for the single-muon trigger, depending on the $\eta$ of the muon, and it is $>$94\% for the single-electron trigger.
In the high-mass \lnujet{} analysis, muons selected online must have $\pt > 45\GeV$ and $\abs{\eta} < 2.1$, while the minimum \pt threshold for electrons is 105\GeV. There are no requirements on the isolation and loose identification criteria are used, since these introduce inefficiencies at high resonance masses.
The selection efficiencies with respect to the offline requirements of the single-muon trigger vary between 90\% and 95\%. The efficiency is above 98\% for the single-electron trigger.

In the dijet channel, events are selected online using a variety of different hadronic triggers based on the scalar \pt sum of all jets in the event ($\HT$) or the presence of at least one jet with loose substructure requirements; the details of jet substructure are described in Section~\ref{subsec:Vhadr}.
Events must satisfy at least one of the following four requirements. The first requirement is simply $\HT > 800\GeV$. The second requirement is $\HT > 650\GeV$ and a difference in $\eta$ between the two leading jets in the event satisfying the condition $\Delta\eta < 1.5$. The accepted jets are further required to have a dijet invariant mass $>$ 900\GeV. The third criterion is that at least one jet with $\pt > 360\GeV$ and a trimmed mass (as defined in Section~\ref{subsec:Vhadr}) $\mJ > 30\GeV$ is present in the event. Fourthly, events with $\HT > 700\GeV$ and at least one jet with $\mJ > 50\GeV$ are also selected for analysis.

The pp data collected by CMS with the detector in its fully operational state correspond to 2.3\fbinv of integrated luminosity~\cite{CMS-PAS-LUM-15-001}. Additional data equivalent of 0.37\fbinv of integrated luminosity were collected with the HF running in suboptimal conditions; those data are used only for the dijet channel, since jets reconstructed online and used for the calculation of \HT{} are in the range of $\abs{\eta} < 3.0$. The trigger efficiency is found to be unaffected by the condition of the HF.

Offline, all events are required to have at least one primary interaction vertex reconstructed within a 24\cm window along the beam axis, with a transverse distance from the mean pp interaction point of less than \unit{2}{\cm}~\cite{Chatrchyan:2014fea}. In the presence of more than one vertex passing these requirements, the primary interaction vertex is chosen to be the one with the highest total $\pt^2$, summed over all the associated tracks.

\subsection{Jet reconstruction}

Jets are clustered from the four-momenta of the particles reconstructed using the CMS PF algorithm, from the \FASTJET software package~\cite{Cacciari:2011ma}.
In the jet clustering procedure charged PF candidates not associated with the primary interaction vertex are excluded.
Jets used for identifying the W and \Zo boson decays to \qqbar are clustered using the anti-\kt algorithm~\cite{Cacciari:2008gp} with distance parameter $R = 0.8$ (``AK8 jets'').
To identify b jets, the anti-\kt jet clustering algorithm is used with $R = 0.4$ (``AK4 jets''), along with the inclusive combined secondary vertex b tagging algorithm~\cite{Chatrchyan:2012jua,CMS-PAS-BTV-15-001}.
The chosen algorithm working point provides a misidentification rate of $\approx$1\% and efficiency of $\approx$70\%.
A correction based on the area of the jet projected on the front face of the calorimeter is used to take into account the extra energy clustered in jets due to neutral particles coming from pileup.
Jet energy corrections are obtained from simulation and from dijet and  photon+jet events in data, as discussed in Ref.~\cite{CMS:JetCalibration}.
Additional quality criteria are applied to the jets to remove spurious jet-like features originating from isolated noise patterns in the calorimeters or the tracker.
The efficiency of these requirements for signal events is above 99\%.
In the \lnujet{} channel, the AK8 and AK4 jets are required to be separated from any well-identified muon or electron by $\Delta R=\sqrt{\smash[b]{(\Delta\phi)^2+(\Delta\eta)^2}}>0.8$ and $>$0.3, respectively.
All AK4 and AK8 jets must have $\pt > 30\GeV$ and $>$200\GeV, respectively, and $\abs{\eta}<2.4$ to be considered in the subsequent steps of the analysis.

\subsection{Final reconstruction and selection of leptons and missing transverse momentum}\label{subsec:leptons}

Muons are reconstructed through a fit to hits in both the inner tracking system and the muon spectrometer~\cite{Chatrchyan:2012xi}.
Muons must satisfy identification requirements on the impact parameters of the track,  the number of hits reconstructed in both the silicon tracker and the muon detectors, and the uncertainty in the \pt.
These quality criteria ensure a precise measurement of the four-momentum and rejection of misreconstructed muons.
An isolation requirement is applied to suppress background from multijet events where jet constituents are identified as muons. A cone of radius $\Delta R= 0.3$ is constructed around the muon direction, and the isolation parameter is defined as the scalar sum of the \pt of all the additional reconstructed tracks within the cone, divided by the muon \pt. The efficiency of this muon selection has been measured through a ``tag-and-probe'' method using \Zo bosons~\cite{CMS:FirstInclZ}, and it has a negligible dependence on the pileup.
In the high-mass \lnujet analysis, events must have exactly one isolated muon with $\pt > 53\GeV$ and $\abs{\eta} < 2.1$. A looser \pt requirement of 40\GeV is used for the low resonance mass range.

Electron candidates are required to have a match between energy deposited in the ECAL and momentum determined from the reconstructed track~\cite{Khachatryan:2015hwa}.
To suppress multijet background, electron candidates must pass stringent identification and isolation criteria~\cite{Chatrchyan:2012meb}.
Those criteria include requirements on the geometrical matching between ECAL depositions and position of reconstructed tracks, the ratio of the energies deposited in the HCAL and ECAL, the distribution of the ECAL depositions, the impact parameters of the track, and the number of reconstructed hits in the silicon tracker.
In the high-mass \lnujet analysis, we require exactly one electron with $\pt > 120\GeV$ and $\abs{\eta} < 2.5$. A looser \pt requirement of 45\GeV is used for the low resonance mass range. Reconstructed electrons must also be located outside of the transition region between the ECAL barrel and endcaps ($1.44 < \abs{\eta} <1.57$), because the reconstruction of an electron object in this region is not optimal.

The missing transverse momentum \PTm is defined as the magnitude of the vector sum of the transverse momenta of the reconstructed PF objects.
The value of \PTm is modified to account for corrections to the energy scale of all the reconstructed AK4 jets in the event.
More details on the \PTm performance in CMS can be found in Refs.~\cite{CMS:METperformances,Khachatryan:2014gga}.
Requirements of $\PTm > 40$ and $> 80\GeV$ are applied, respectively, in the muon and electron channels in the \lnujet analysis.
The threshold is higher in the electron channel to further suppress the larger background from multijet processes.
Since the \PTm calculation requires the detector to provide complete geometric coverage,
events in data without fully operational HF calorimeter are not considered for the \lnujet channel.

\subsection{The identification of \texorpdfstring{$\PW$/$\cPZ \to \Pq\Paq$}{W to q anti-q' and Z to q anti-q} using jet substructure}
\label{subsec:Vhadr}

The AK8 jets are used to reconstruct the W jet and Z jet candidates from their decays to highly boosted quark jets. To discriminate against multijet backgrounds, we exploit both the reconstructed jet mass, which is required to be close to the W or \Zo boson mass, and the two-prong jet substructure produced by the particle cascades of two high-\pt quarks that merge into one jet~\cite{Khachatryan:2014vla}. Jets that are identified as arising from the merged decay products of a V boson are hereafter referred to as ``V jets''.

As a first step in exploring potential substructure, the jet constituents are subjected to a jet grooming algorithm that improves the resolution in the jet mass and reduces the effect of pileup~\cite{Chatrchyan:2013vbb}.
The goal of the algorithm is to recluster the jet constituents, while applying additional requirements that eliminate soft, large-angle QCD radiation
that increases the jet mass relative to the initial V boson mass. Different jet grooming algorithms have been explored at CMS, and their performance on jets in multijet processes has been studied in detail~\cite{Chatrchyan:2013vbb}.
In this analysis, we use the \textit{jet pruning}~\cite{jetpruning1,Ellis:2009me} algorithm for the main analysis and the \textit{jet trimming} algorithm~\cite{Krohn:2009th} at the trigger level as well as for cross checks.
Jet pruning reclusters each AK8 jet starting from all its original constituents, through the implementation of the Cambridge-Aachen (CA) algorithm~\cite{Catani:1993hr,Dokshitzer:1997in} to discard ``soft'' recombinations in each step of the iterative CA procedure.
The pruned jet mass, \mJ{}, is computed from the sum of the four-momenta of the constituents that are not removed by the pruning; it is then scaled by the same factor as that used to correct the original jet \pt. The jet is considered as a V jet candidate if \mJ{} falls in the range $\SRLOW < \mJ < \SRHIGH\GeV$, which we define as the signal jet mass window. In the low-mass analysis, only W jet candidates are considered, thus the mass window applied is $65 < \mJ < 95\GeV$.

Additional discrimination against jets from gluon and single-quark hadronization is obtained from the quantity called \textit{N-subjettiness}~\cite{Thaler:2010tr}. The constituents of the jet before the pruning procedure are reclustered using the \kt algorithm~\cite{Catani:1993hr,Ellis:1993tq}, until $N$ joint objects (\textit{subjets}) remain in the iterative combination procedure of the algorithm. The $N$-subjettiness, $\tau_{N}$, is then defined as
\begin{equation}
\tau_N = \frac{1}{d_{0}} \sum_{k} p_{\mathrm{T},k} \min( \Delta R_{1,k}, \Delta R_{2,k},\ldots,\Delta R_{N,k}),
\end{equation}
where the index $k$ runs over the PF constituents of the jet, and the distances $\Delta R_{n,k}$ are calculated relative to the axis of the $n$-th subjet.
The normalization factor $d_{0}$ is calculated as $d_{0}=\sum_{k} p_{\mathrm{T},k}R_{0}$, setting $R_{0}$ to the distance parameter used in the clustering of the original jet.
The variable $\tau_{N}$ quantifies the compatibility of the jet clustering with the hypothesis that exactly $N$ subjets are present, with small values of $\tau_{N}$ indicating greater compatibility.
The ratio between 2-subjettiness and 1-subjettiness, $\nsubj=\tau_{2}/\tau_{1}$, is found to be a powerful discriminant between jets originating from hadronic V decays and from gluon and single-quark hadronization.
Jets from W or \Zo decays in signal events are characterized by lower values of $\nsubj$ relative to SM backgrounds.
We reject V jet candidates with $\nsubj>0.75$. The remaining events are further categorized according to their value of \nsubj{} to enhance the sensitivity of the analysis, as summarized in Table~\ref{tab:WtaggingScaleFactors}.

Since data/simulation discrepancies in the jet substructure variables $\mJ$ and $\nsubj$ can bias the signal
efficiency estimated from simulated samples, the modelling of signal efficiency is cross-checked in a signal-free
sample with jets having characteristics that are similar to those expected for a genuine signal.
A sample of high-\pt W bosons that decay to quarks, and are reconstructed as single AK8 jets, is studied in \ttbar and single top quark events.
Scale factors for the $\nsubj{}$ selection efficiency are extracted following Ref.~\cite{Khachatryan:2014vla}.
In this method, the pruned jet mass distributions of events that pass and fail the $\nsubj$ selection are fitted simultaneously to separate the W boson signal from the combinatorial components in the top quark enriched sample in both data and simulation.
The scale factors are listed in Table~\ref{tab:WtaggingScaleFactors} and are used to correct the total signal efficiency and the VV background normalization predicted by the simulation.
The uncertainties in the scale factors quoted for the $\nsubj$ selection include two systematic uncertainties. One comes from the modelling of the nearby jets and \pt spectrum in \ttbar MC events, obtained by comparing LO and NLO \ttbar simulation. The other is due to the choice of the models used to fit signal and background. The quadratic sum of these systematic uncertainties is found to be smaller than half of the statistical uncertainty in the scale factor.
An additional uncertainty is calculated to account for the extrapolation of the scale factor from \ttbar events with an average jet $\pt\approx200\GeV$ to higher momenta. This is estimated from the difference between \PYTHIA{} and \HERWIG{++}~\cite{Bahr:2008pv} showering  models with resulting factors of
$4.5\% \ln (\pt/200\GeV)$
and $5.9\%  \ln (\pt/200\GeV)$
for $\tau_{21} < 0.6$ and $\tau_{21} < 0.45$, respectively.
For the $0.45<\tau_{21}<0.75$ selection, this uncertainty is increased by the ratio of the uncertainties in the scale factors shown in Table~\ref{tab:WtaggingScaleFactors} ($0.32/0.06$),
and treated as anti-correlated with the uncertainty for $\tau_{21}<0.45$.
The mean $\langle\mJ\rangle$ and resolution $\sigma$ value of the Gaussian component of the fitted W jet mass are also extracted to obtain corrections that are applied to the simulated pruned jet mass.
The values are listed in Table~\ref{tab:Wmass}, where the quoted uncertainties are statistical.
The mass peak position is slightly shifted relative to the W boson mass because of the extra energy deposited in the jet cone from pileup, underlying event, and initial-state radiation not completely removed in the jet pruning procedure. For events with top quarks, additional energy contributions arise also from the possible presence of a b jet close to the W jet candidate.
Because the kinematic properties of W jets and Z jets are very similar, the same corrections are also used when the V jet is assumed to arise from a \Zo boson.

\begin{table}[htbp]
   \centering
   \topcaption{Data-to-simulation scale factors for the efficiency of the \nsubj{} selection used in the analyses, as extracted from top quark enriched data and from simulation.}
   \begin{tabular}{rc}
   \hline
   \nsubj{} selection & Efficiency scale factor\\
   \hline
   $\nsubj < 0.45$              & \SFWTAGHPWPT \\
   $0.45 < \nsubj < 0.75$       & \SFWTAGLPWPT \\
   \hline
   $\nsubj < 0.6 \, \, \,$             & \SFWTAGHPWPL \\
   \hline
   \end{tabular}
   \label{tab:WtaggingScaleFactors}
\end{table}

\begin{table}[htbp]
   \centering
   \topcaption{The W jet mass peak position and resolution, as extracted from top quark enriched data and from simulation. These results are used to apply corrections in the V tagging procedure.}
   \begin{tabular}{lcc}
   \hline
   $\nsubj < 0.45$ & $\langle\mJ\rangle$ (\GeVns{}) & $\sigma$ (\GeVns{})\\
   \hline
   Data          & \WMASSDATAWPT & \WRESDATAWPT\\
   Simulation & \WMASSMCWPT    & \WRESMCWPT\\
   \hline
   \end{tabular}
   \label{tab:Wmass}
\end{table}

\subsection{The reconstruction and identification of \texorpdfstring{$\PW\to \ell \nu$}{W to l nu}}
\label{subsec:Vlept}

In the \lnujet channel, identified muons and electrons are associated with $\PW \to \ell\Pgn$ candidates. The \pt of the undetected neutrino is assumed to be equal to the \PTm. The longitudinal momentum of the neutrino ($p_{z}$) is obtained by solving a quadratic equation that sets the $\ell\Pgn$ invariant mass to the known W boson mass~\cite{Agashe:2014kda}. In the case of two real solutions, we choose the one with smaller $p_{z}$; in the case of two complex solutions, we use their real part. The four-momentum of the neutrino is used to reconstruct the four-momentum of the $\PW \to \ell \nu$ candidate.

\subsection{Final event selection and categorization}
\label{sec:finalSelection}

After reconstructing the two vector bosons, we apply the final criteria in the search. For all channels, any V boson candidate is required to have $\pt > 200$\GeV. In addition, there are specific selection criteria chosen for the \lnujet{} and dijet analyses. For the \lnujet channel, we reject events with more than one well-identified muon or electron.
We also require that the two V bosons from the decay of a massive resonance are approximately back-to-back: the $\Delta R$ between the lepton and the V jet is greater than 1.6;
the $\Delta\phi$ between the vector \ptvecmiss and the W jet, as well as between the $\PW \to \ell \nu$ and V jet candidates, are both greater than 2 radians.
To further reduce the level of the \ttbar background in the \lnujet channel, events are rejected if they contain one or more b-tagged AK4 jets. This veto preserves about 90\% of the signal events.
For the dijet analysis, we require the two AK8 jets to have $\abs{\Delta\eta_{jj}} < 1.3$, while the dijet system invariant mass \mjj{} must be above 1\TeV.

To enhance the analysis sensitivity, events are categorized according to the characteristics of the V jet.
For the dijet and high-mass \lnujet channels, the V jet is deemed a W or Z boson candidate if its pruned mass falls in the range
\unit{\SRLOW--\SRMIDDLE} or \unit{\SRMIDDLE--\SRHIGH}{\GeV}.
This leads to three categories for the dijet channel (WW, WZ, and ZZ), and two categories for the \lnujet{} channel (WW and WZ).
For the low-mass \lnujet channel, only W jets are considered in the signal region $65 < \mJ < 95\GeV$.
In addition, in the low- and high-mass \lnujet channels, V jets are selected to have $\tau_{21} \leq 0.45$ and $\leq$0.6, respectively.
A tighter selection is required for the low-mass analysis as more background is expected in that mass range.
In the dijet channel, we select ``high-purity" (HP) and ``low-purity" (LP) V jets by requiring $\tau_{21} \leq 0.45$ and $ 0.45 <\tau_{21} < 0.75$, respectively.
Events are always required to have one HP V jet, and are divided into HP or LP events, depending on whether the other V jet is of high or low purity.
Although the HP category dominates the total sensitivity of the analysis, the LP category is retained since for heavy resonances it can improve the signal efficiency with only moderate background contamination. The final categorization is therefore based on two and four classes of events for the low- and high-mass \lnujet channels, respectively, depending on their lepton flavor (muon or electron), and V jet mass category (W or Z). For the dijet analysis, categorization in V jet purity and mass category (WW, WZ, and ZZ) yields a total of 6 orthogonal classes of events.

The two boson candidates are combined into a diboson candidate, with presence of signal then inferred from the observation of localized excesses in the \mVV distribution.
When several diboson resonance candidates are present in the same event, only the one with the highest \pt V jet (\lnujet{} analyses) or the two highest mass V jets (dijet analysis) are retained.

\par A summary of the final event selections and categories is presented in Table~\ref{tab:cutsummary_lvJ} for the \lnujet{} analyses and in
Table~\ref{tab:cutsummary_JJ} for the dijet analysis.

\begin{table}[!htb]
\centering
\topcaption{Summary of the final selections and categories for the \lnujet{} channel. The values indicated in parentheses correspond to the low-mass analysis.}
\label{tab:cutsummary_lvJ}
\renewcommand{\arraystretch}{1.2}
\begin{tabular}{lc}
\hline
\multicolumn{1}{c}{Selection} & Value\\
\hline \hline
\multicolumn{1}{c}{Lepton selections}\\
\cline{1-1}
Electron & $\pt > 120~(45)\GeV$\\
              & $\abs{\eta} < 2.5$ (except $1.44 < \abs{\eta} <1.57$)\\
Muon     & $\pt > 53~(40)\GeV$\\
             & $\abs{\eta} < 2.1$\\
Number of electrons & exactly 1\\
Number of muons & exactly 1\\
\hline
\multicolumn{1}{c}{AK4 jet selections}\\
\cline{1-1}
Jet $\pt$ &  $\pt > 30\GeV$\\
Jet $\eta$  & $\abs{\eta} < 2.4$\\
Number of b-tagged AK4 jets & 0\\
\hline
\multicolumn{1}{c}{\PTm selections}\\
\cline{1-1}
\PTm (electron channel) & $\PTm  > 80\GeV$\\
\PTm (muon channel) & $\PTm  > 40\GeV$\\
\hline
\multicolumn{1}{c}{Boson selections}\\
\cline{1-1}
W $\to\ell\nu$      &  $\pt > 200\GeV$\\
V $\to\qqbar$ (AK8 jet)     &  $\pt > 200\GeV$\\
                               & $\abs{\eta} < 2.4$\\
Back-to-back topology & $\Delta R (\ell , \mathrm{V}_{\qqbar}) > 1.6$\\
                      & $\Delta \phi (\mathrm{V}_{\qqbar} , \mbox{\PTm})>2$\\
                      & $\Delta \phi (\mathrm{V}_{\qqbar} , \mathrm{W}_{\ell\nu})>2$\\
Pruned jet mass        & $65 < \mJ < 105~(95) \GeV$\\
2- to 1-subjettiness ratio & $\nsubj < 0.60~(0.45)$\\
\hline
\multicolumn{1}{c}{\mJ{} categories (only for high-mass analysis)}\\
\cline{1-1}
WW   & $ 65 < \mJ < 85\GeV$\\
WZ   & $ 85 < \mJ < 105\GeV$\\
\hline						
\end{tabular}
\end{table}

\begin{table}[!htb]
\centering
\topcaption{Summary of the final selections and categories for the dijet analyses.}
\label{tab:cutsummary_JJ}
\renewcommand{\arraystretch}{1.2}
\begin{tabular}{lc}
\hline
\multicolumn{1}{c}{Selection} & Value\\
\hline \hline
\multicolumn{1}{c}{Boson selections}\\
\cline{1-1}
V $\to\qqbar$ (2 AK8 jets) & $\pt >200\GeV$\\
  & $\abs{\eta} < 2.4$\\
Pruned jet mass & $65 < {\mJ}_1,{\mJ}_2  < 105\GeV$\\
Topology    & $|\Delta \eta_\mathrm{jj}| < 1.3$\\
Dijet invariant mass     & $\mjj >1\TeV$\\
2- to 1-subjettiness ratio    & $\nsubj < 0.75$\\
\hline
\multicolumn{1}{c}{\mJ{} categories}\\
\cline{1-1}
WW & $ 65 < {\mJ}_{1} < 85\GeV$, $ 65 < {\mJ}_{2} < 85\GeV$\\
WZ & $ 65 < {\mJ}_{1} < 85\GeV$, $ 85 < {\mJ}_{2} < 105\GeV$\\
ZZ & $ 85 < {\mJ}_{1} < 105\GeV$, $ 85 < {\mJ}_{2} < 105\GeV$\\
\hline
\multicolumn{1}{c}{\nsubj{} categories}\\
\cline{1-1}
High-purity   & $\tau_{\text{21, jet1}} < 0.45$, $\tau_{\text{21, jet2}} < 0.45$\\
Low-purity    & $\tau_{\text{21, jet1}} < 0.45$, $0.45 < \tau_{\text{21, jet2}} < 0.75$\\
\hline						
\end{tabular}
\end{table}

\section{Modeling of background and signal}
\label{sec:backgroundestimation}

The $\mVV$ distribution observed in data is dominated by SM background processes where single quark or gluon jets are falsely identified as V jets. Depending on the final state, the dominant processes are multijets (dijet channel) and inclusive W boson production (\lnujet channel). Subdominant backgrounds include \ttbar, single top quark, and nonresonant diboson SM production.

\subsection{Multijet background}
\label{sec:dijetmethod}

In the \lnujet channel, the multijet background is predicted to be negligible from MC simulation, whereas it represents the major contribution in the dijet analysis.
For the latter, we assume that the SM background can be described by a smooth, parametrizable, monotonically decreasing distribution.
The search is performed by separately fitting the background function to each search region
and simultaneously adding resonant Breit--Wigner (BW) forms across all search regions to represent the signal.
The background probability function is defined by empirical functional forms of 3 and 2 parameters, respectively:
\begin{align}
\label{eq:dijet1}
\frac{\rd N}{\rd\mjj}= \frac{ P_0(1-\mjj/\sqrt{s})^{P_1} } { (\mjj/\sqrt{s})^{P_2} }&&\text{and}
&&
\frac{\rd N}{\rd\mjj}= \frac{ P_0 } { (\mjj/\sqrt{s})^{P_2} },
\end{align}
where
\mjj{} is the dijet invariant mass (equivalent to the diboson candidate mass \mVV{} for this channel),
$\sqrt{s}$ is the pp collision energy in the centre of mass,
$P_0$ is a normalization parameter,
and $P_1$ and $P_2$ parametrize the shape of the \mVV{} distribution.
Starting from the two-parameter functional form, a Fisher F-test is used to check at 10\% confidence level (CL)
if additional parameters are needed to model the background distribution. For the WW
categories and the WZ HP category, the two-parameter form is found to describe the data
spectrum sufficiently well, while for all other channels the three-parameter functional form is preferable.
Alternative parametrizations and functions with up to five parameters are also studied as a cross-check.

\begin{figure}[htbp]
\centering
\includegraphics[width=0.85\cmsFigWidth]{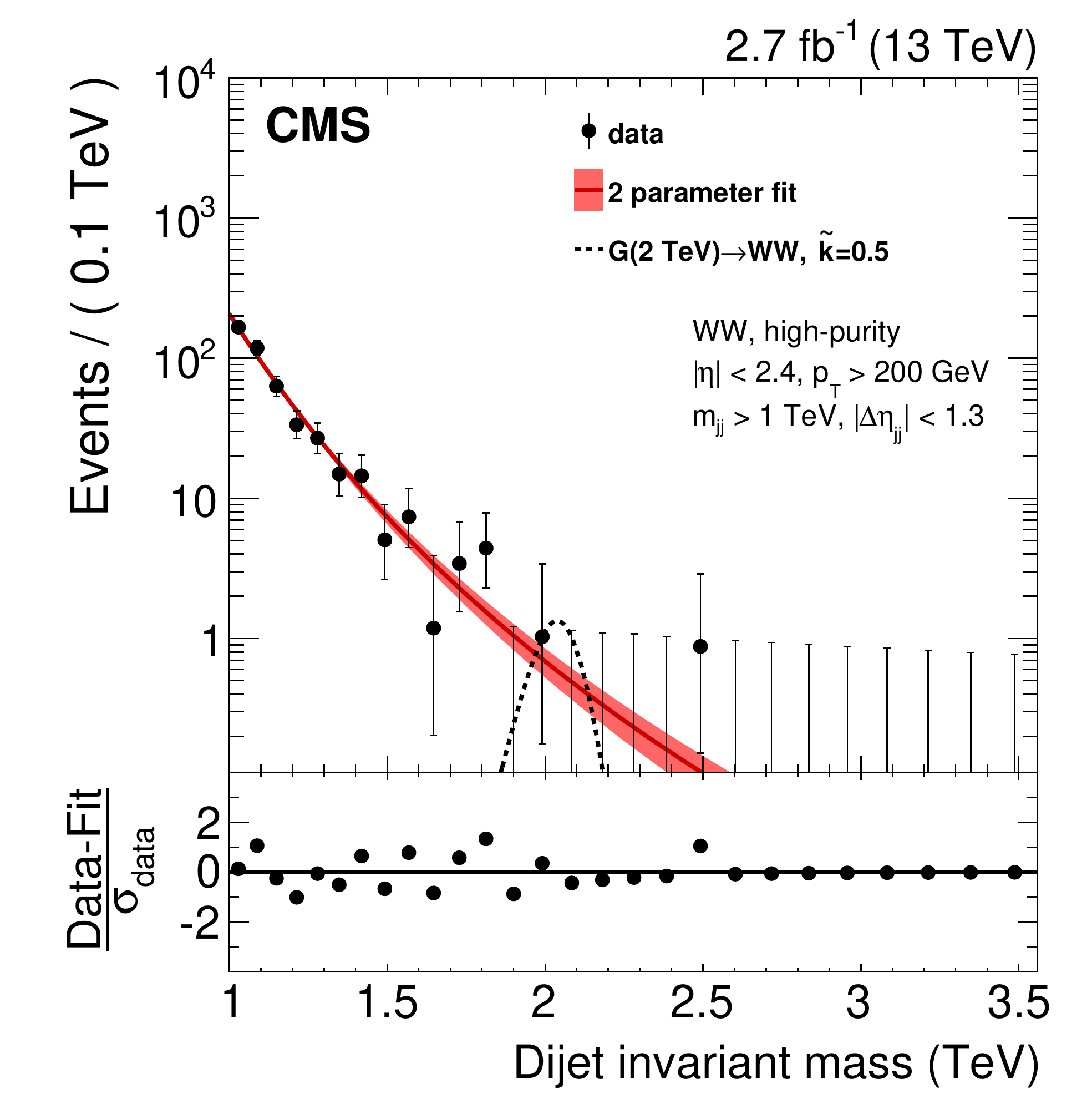}
\includegraphics[width=0.85\cmsFigWidth]{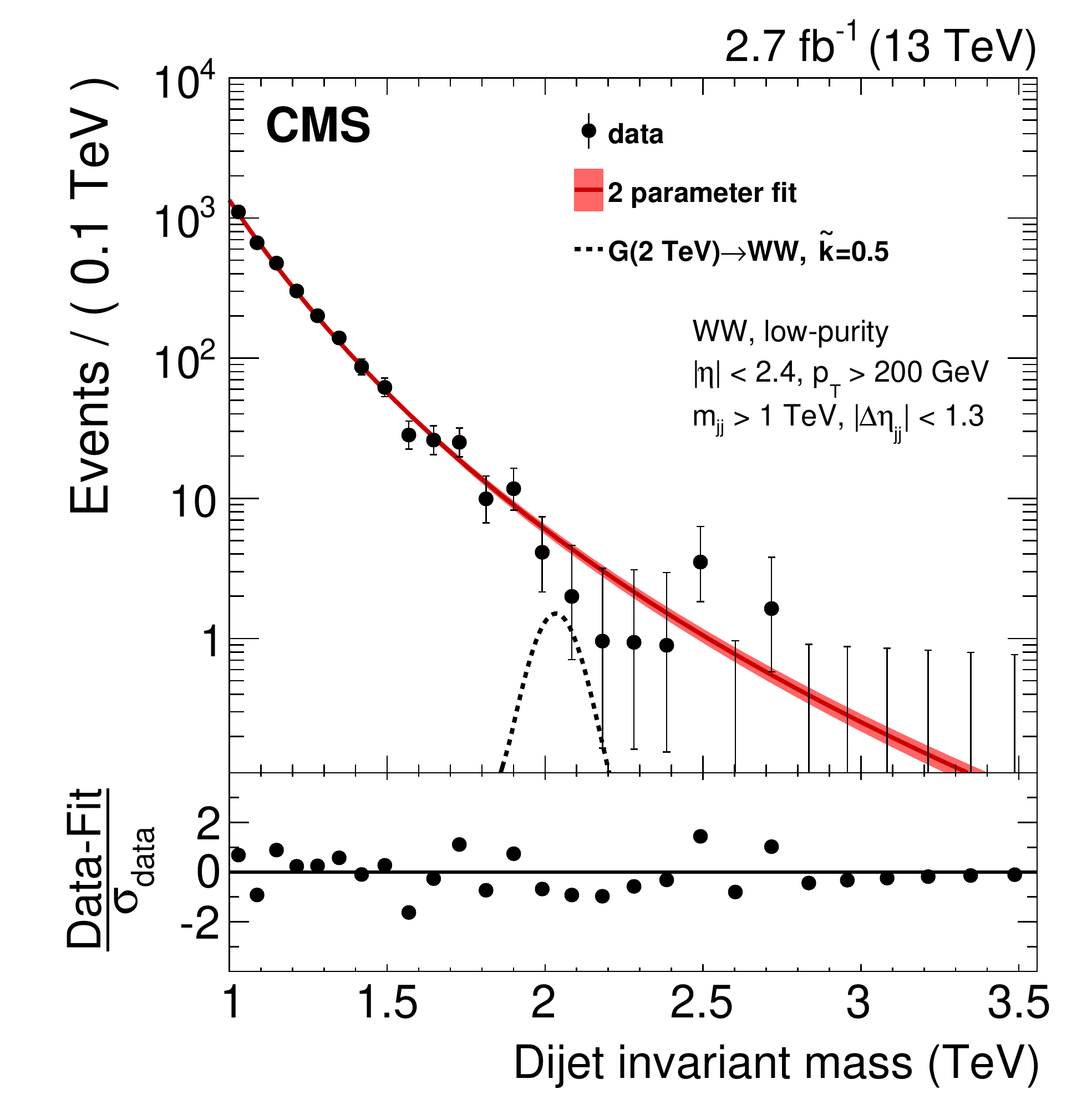}\\
\includegraphics[width=0.85\cmsFigWidth]{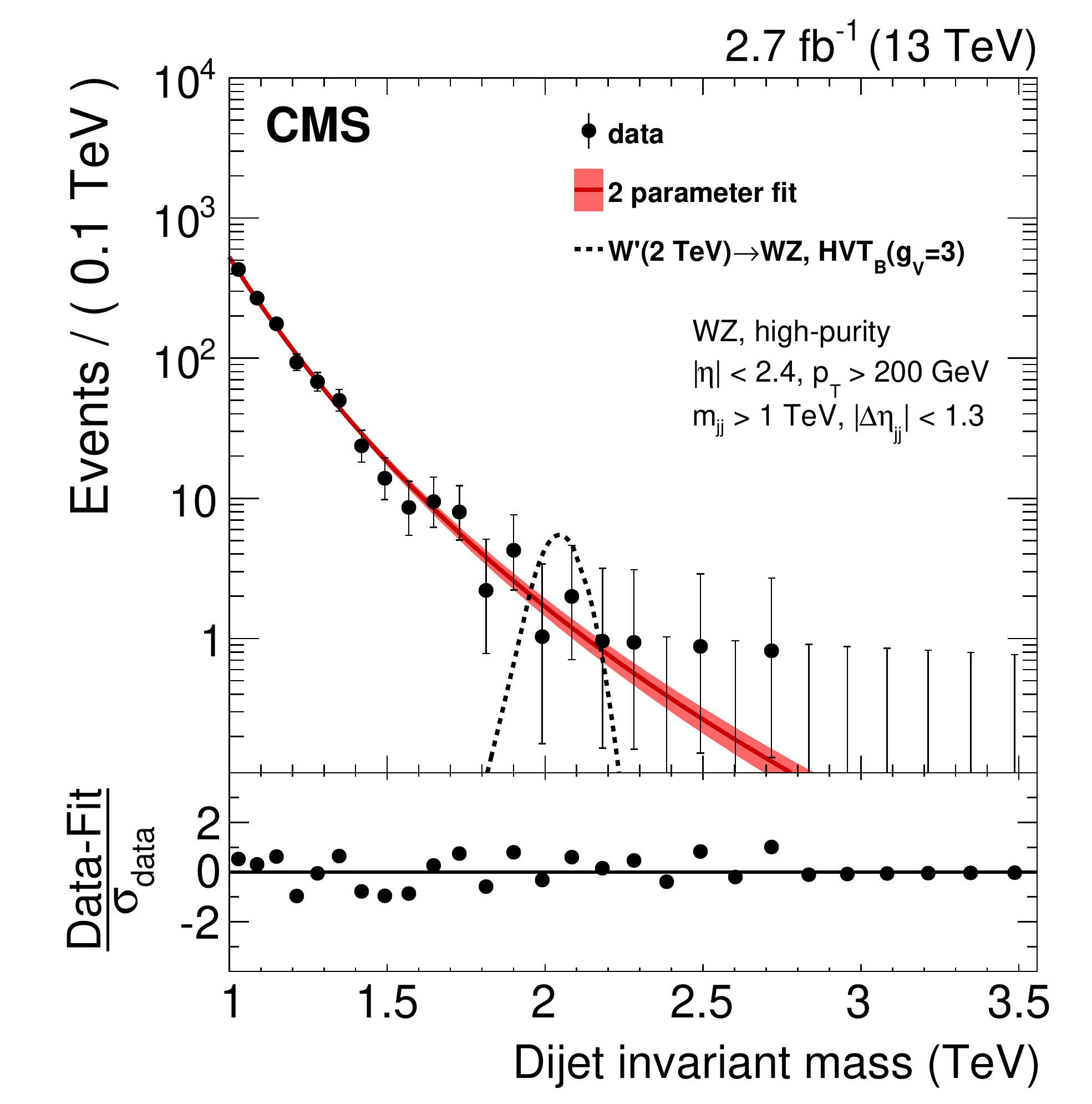}
\includegraphics[width=0.85\cmsFigWidth]{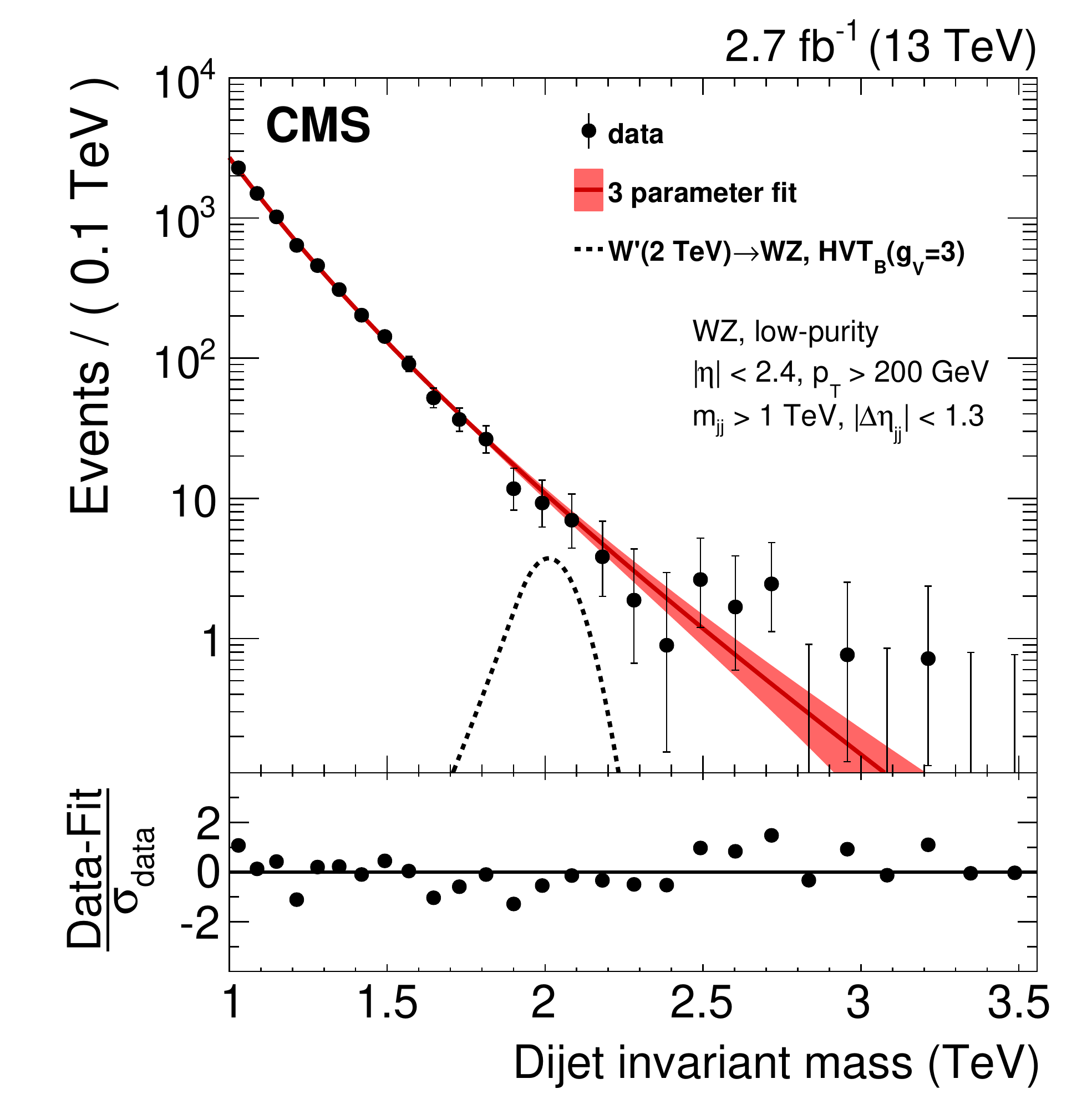}\\
\includegraphics[width=0.85\cmsFigWidth]{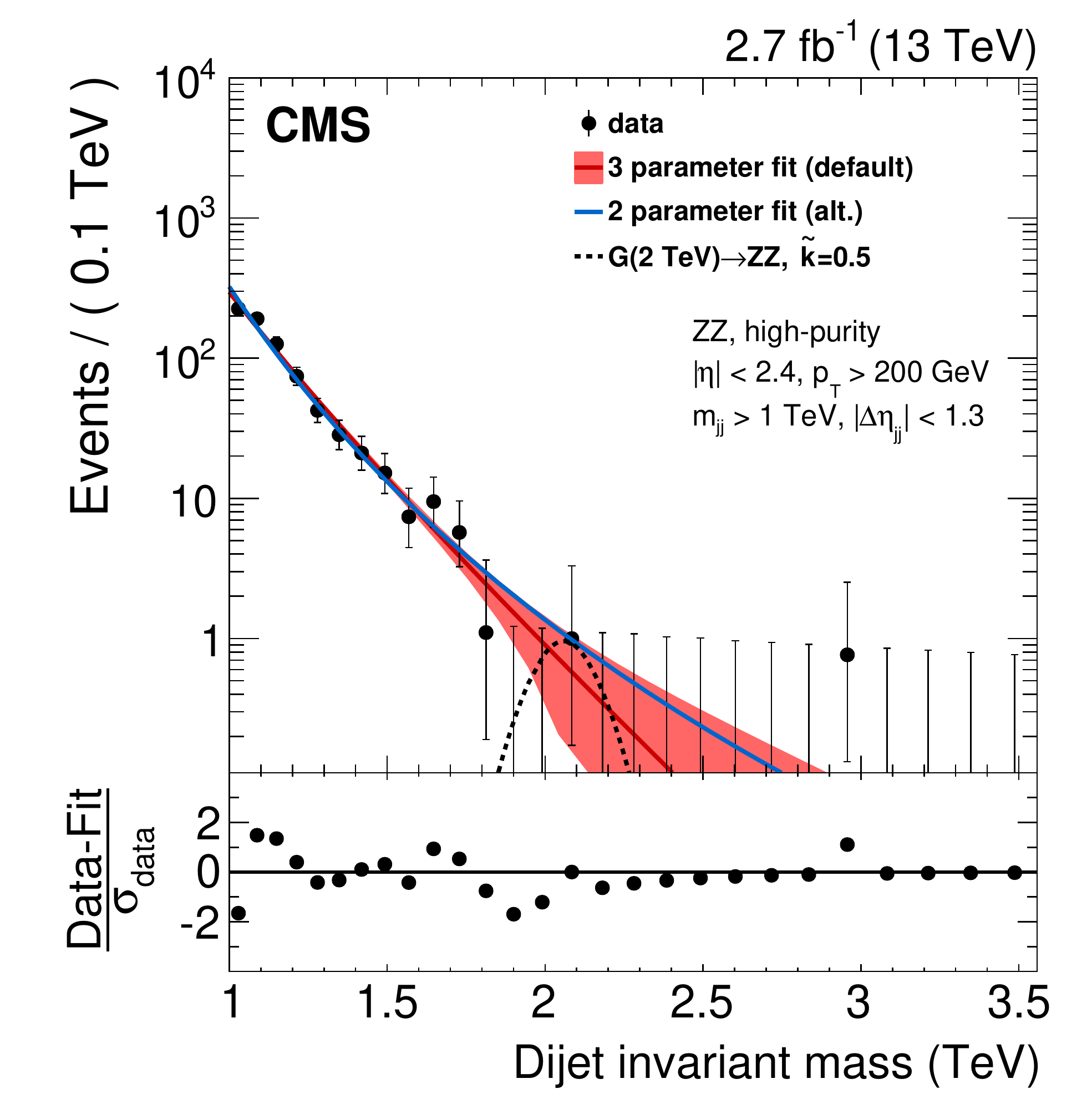}
\includegraphics[width=0.85\cmsFigWidth]{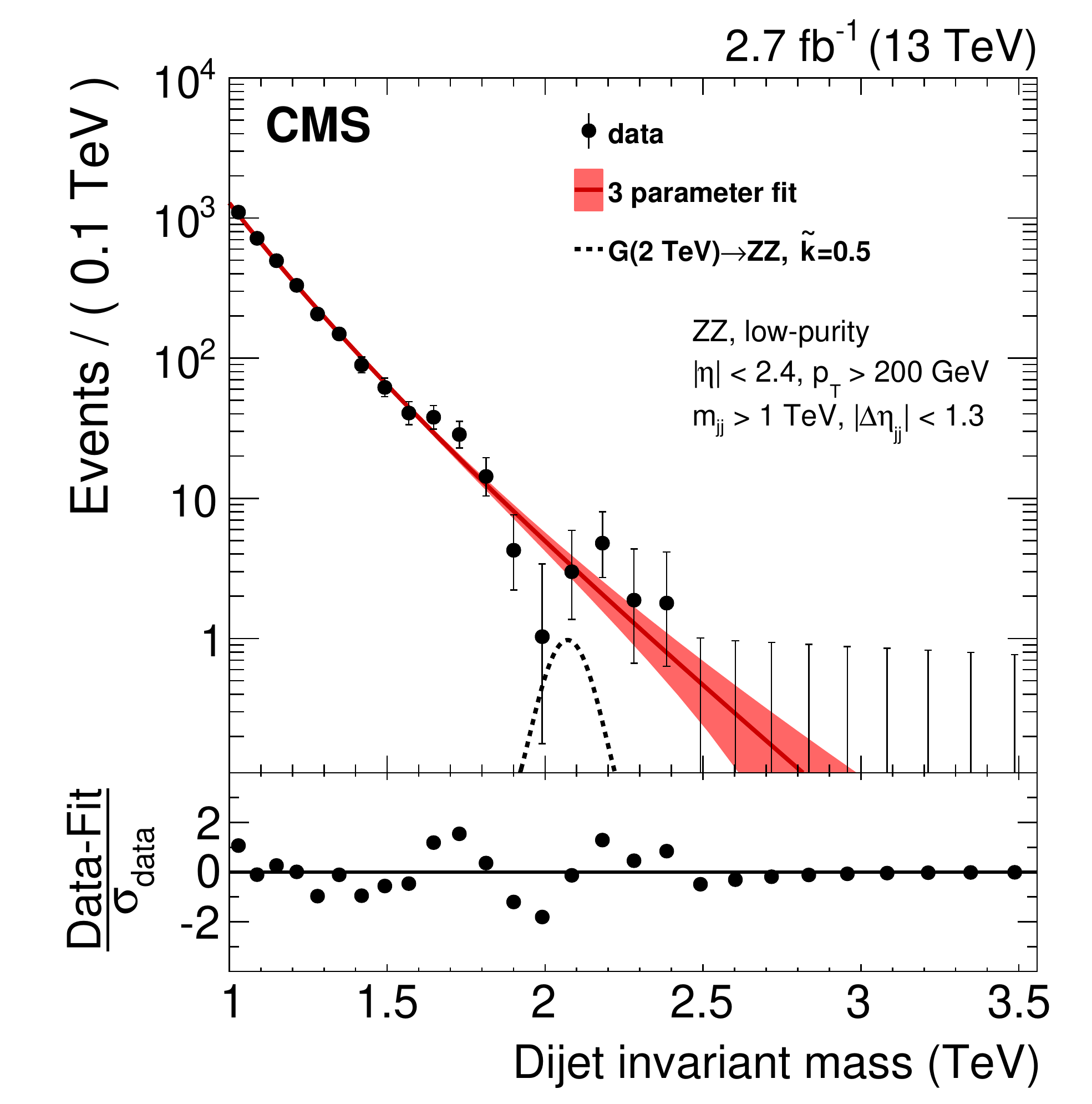}\\
\caption{Final \mjj{} distributions for the dijet analysis in six signal regions.
The high-purity (on the left) and the low-purity (on the right) categories are shown for the WW (top row), WZ (central row), and ZZ (bottom row) \mJ{} regions. The solid curve represents a background-only fit to the data distribution, where the filled red area corresponds to the $\pm$1 standard deviation statistical uncertainties of the fit. The data are represented by the black points. For the ZZ high-purity category (bottom left), we also show the background-only fit using the two-parameter functional form (blue solid line), for comparison.
Signal benchmarks for a mass of 2\TeV are also shown with black dashed lines.
In the lower panel of each plot, the bin-by-bin fit residuals, $(N_\text{data} - N_\text{fit})/\sigma_\text{data}$, are shown.
}
\label{fig:mjjwithFit}
\end{figure}

The binning chosen for the fit reflects the detector resolution.
The fit range is chosen to start where the trigger efficiency reaches its plateau, as this minimizes bias from trigger inefficiency, and to extend to the bin after the highest \mjj{} mass point.
The results are shown in Fig.~\ref{fig:mjjwithFit}.
The solid curve represents the maximum likelihood fit to the data, fixing the number of expected signal events to zero, while the bottom panels show the corresponding pull distributions, quantifying the agreement between the background-only fit and the data.
The expected contributions from bulk graviton and \Wpr resonances with a mass of 2\TeV, scaled to their corresponding cross sections, are given by the dashed curves.

\subsection{Top quark production}
\label{sec:ttbarbackground}

\begin{figure}[!htb]
\centering
 \includegraphics[width=\cmsFigWidth]{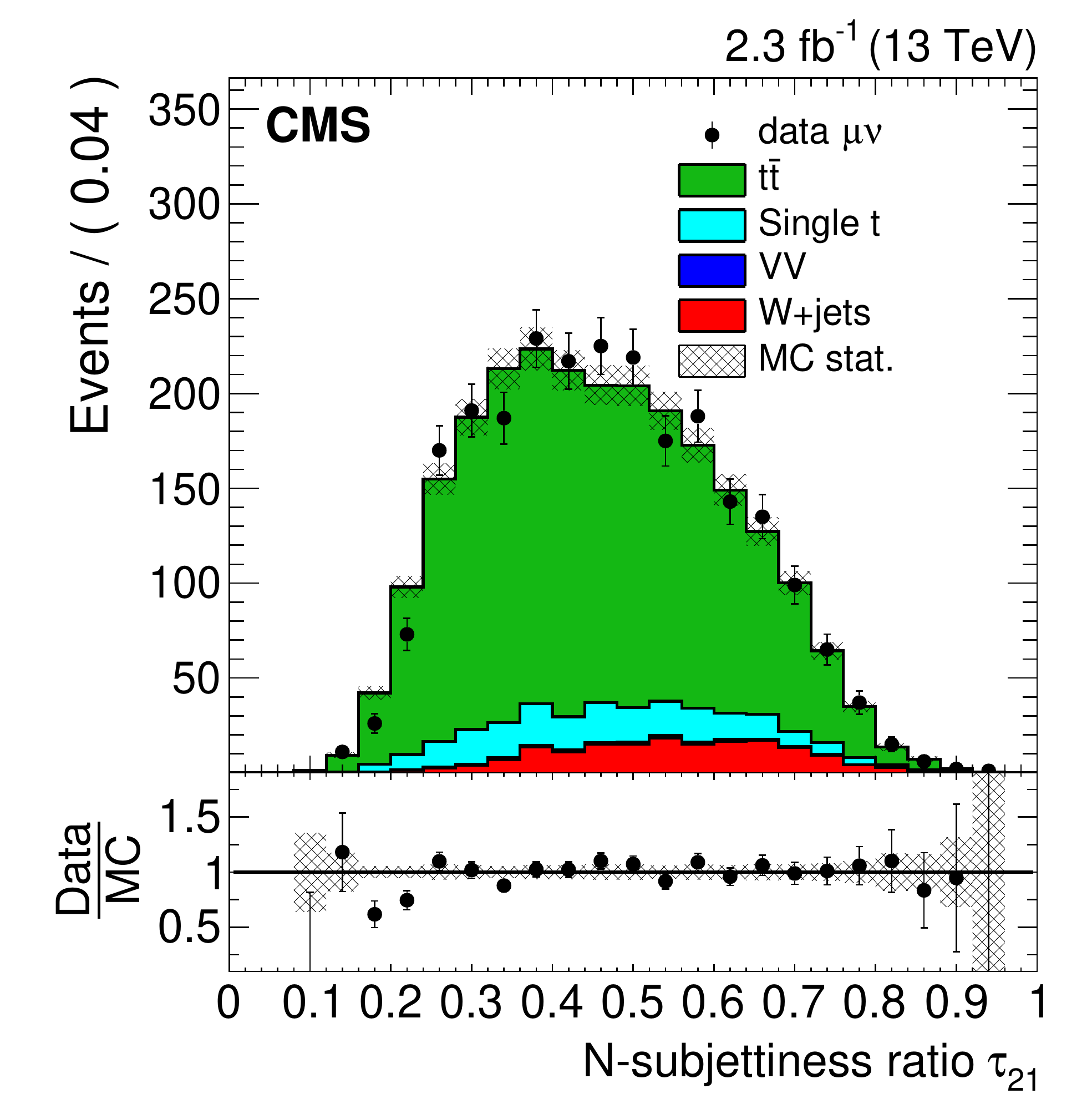}
 \includegraphics[width=\cmsFigWidth]{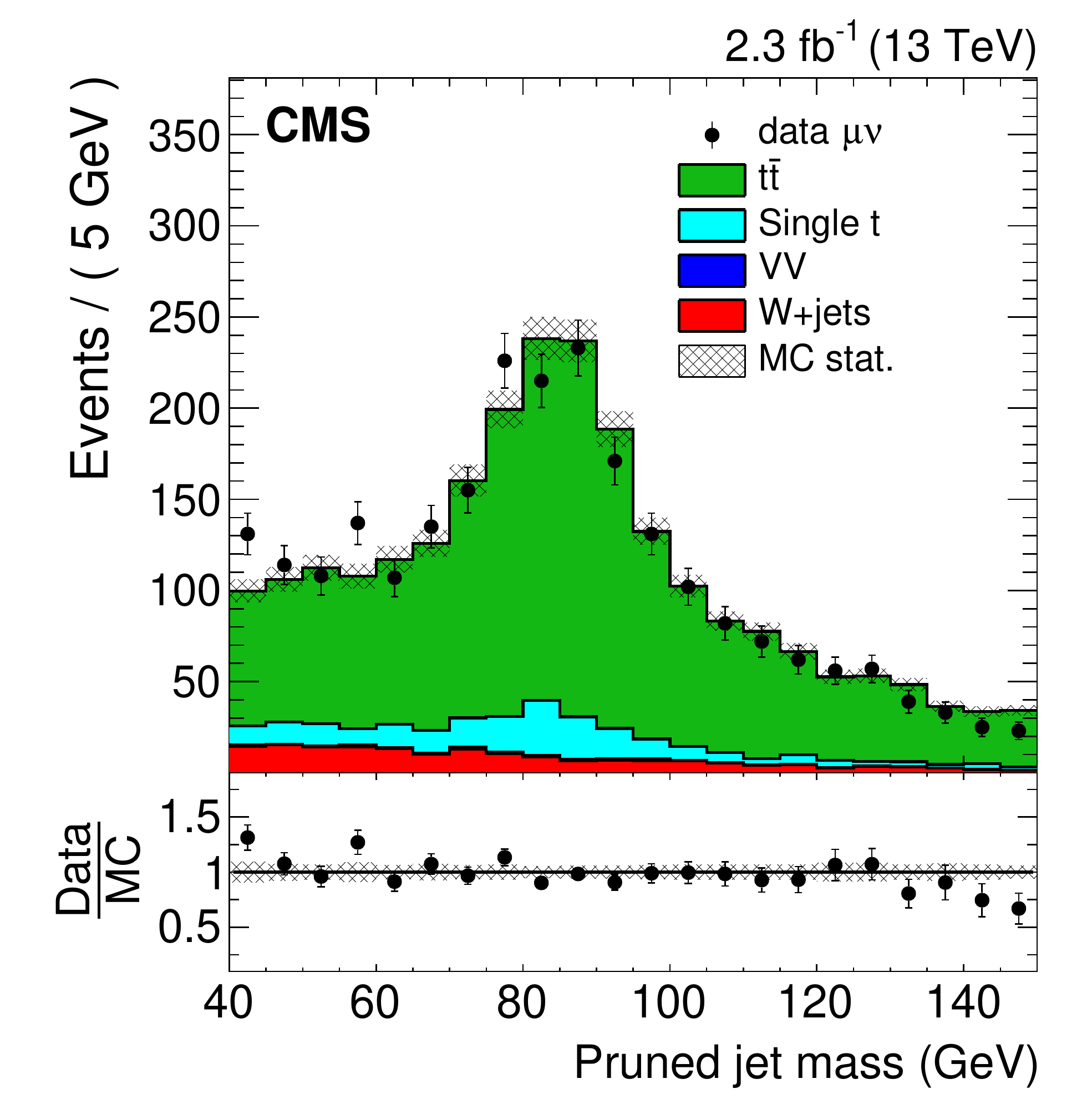}
\caption{Distributions in $N$-subjettiness ratio $\nsubj$ (left) and pruned  $\mJ$ (right) from the top quark enriched control sample in the muon channel.
The \ttbar background is rescaled such that the total number of background events matches the number of events in data.
In the lower panel of each plot, the ratio between data and simulation is shown together with the statistical uncertainty in the simulation normalized by its central value.
}
\label{fig:ttbarControlCut}
\end{figure}

The backgrounds from \ttbar and single top quark production in the \lnujet{} channel are estimated from data-based correction factors in the normalization of the simulation.
A top quark enriched control sample is selected by applying all the analysis requirements in \lnujet{} events
except that the b jet veto is inverted by requiring, instead, at least one b-tagged AK4 jet in the event.
From the comparison between data and simulation, normalization correction factors for \ttbar and single top quark background processes are evaluated in the pruned jet mass regions $65 < \mJ < 105\GeV$ and $65 < \mJ < 95\GeV$, for the electron and muon channels, and for the low- and high-mass selections, separately.
The scale factors, summarized in Table~\ref{tab:topScaleFactors}, include both the W boson signal and the combinatorial components mainly due to events where the extra b jet from the top quark decay is in the proximity of the W, and are used to correct the normalization of the \ttbar and single top quark simulated background predictions in the signal regions.
The $\mJ$ distribution in the top quark enriched sample is shown in the right plot of Fig.~\ref{fig:ttbarControlCut}, while the left plot shows the $\nsubj$ distribution. The \mJ{} distribution shows a clear peak for events with a W boson decaying to hadrons, including the combinatorial background.

\begin{table}[!htb]
   \centering
   \topcaption{Data-to-simulation scale factors for \ttbar and single top quark background processes, extracted from the comparison between
data and simulation in the top quark enriched control sample.}
   \begin{tabular}{lcc}
   \hline
   \nsubj{} selection & Muon channel & Electron channel \\
   \hline
   $\nsubj < 0.6$ (high-mass)             & $0.87 \pm 0.04$ & $0.83 \pm 0.07$ \\
   $\nsubj < 0.45$ (low-mass)             & $0.85 \pm 0.05$ & $0.86 \pm 0.08$ \\
   \hline
   \end{tabular}
   \label{tab:topScaleFactors}
\end{table}

\subsection{The W+jets background}
\label{sec:alphamethod}

The W+jets background in the \lnujet channel is estimated through the \emph{$\alpha$ ratio method}.
This method assumes that the correlation between \mJ{} and \mVV{} for the dominant W+jets background can be adequately modelled by simulation.
A signal-depleted control region (sideband) is defined by requiring the mass of the V jet to lie below or above the nominal selection; the \mVV{} distribution observed in this region is then extrapolated to the nominal region through a transfer function estimated from simulation. Other minor sources of background, such as \ttbar, single top quark, and SM diboson production, are estimated from simulation after applying correction factors based on control regions in data, as described in Sections~\ref{subsec:Vhadr} and \ref{sec:ttbarbackground}. The sideband region is defined around the jet mass window described in Section~\ref{sec:eventreconstruction}. The lower and upper sidebands correspond to the \mJ{} ranges \LOWSBLOW--\SRLOW and \HIGHSBLOW--\HIGHSBHIGH{\GeV}, respectively. The Higgs boson mass region, defined by the range \SRHIGH--\HIGHSBLOW{\GeV}, corresponds to the signal region of searches for diboson in final states with highly Lorentz-boosted Higgs bosons~\cite{Khachatryan:2016cfx}, and is therefore not used to estimate the background.

\par
\begin{figure}[htbp]
\centering
\includegraphics[width=\cmsFigWidth]{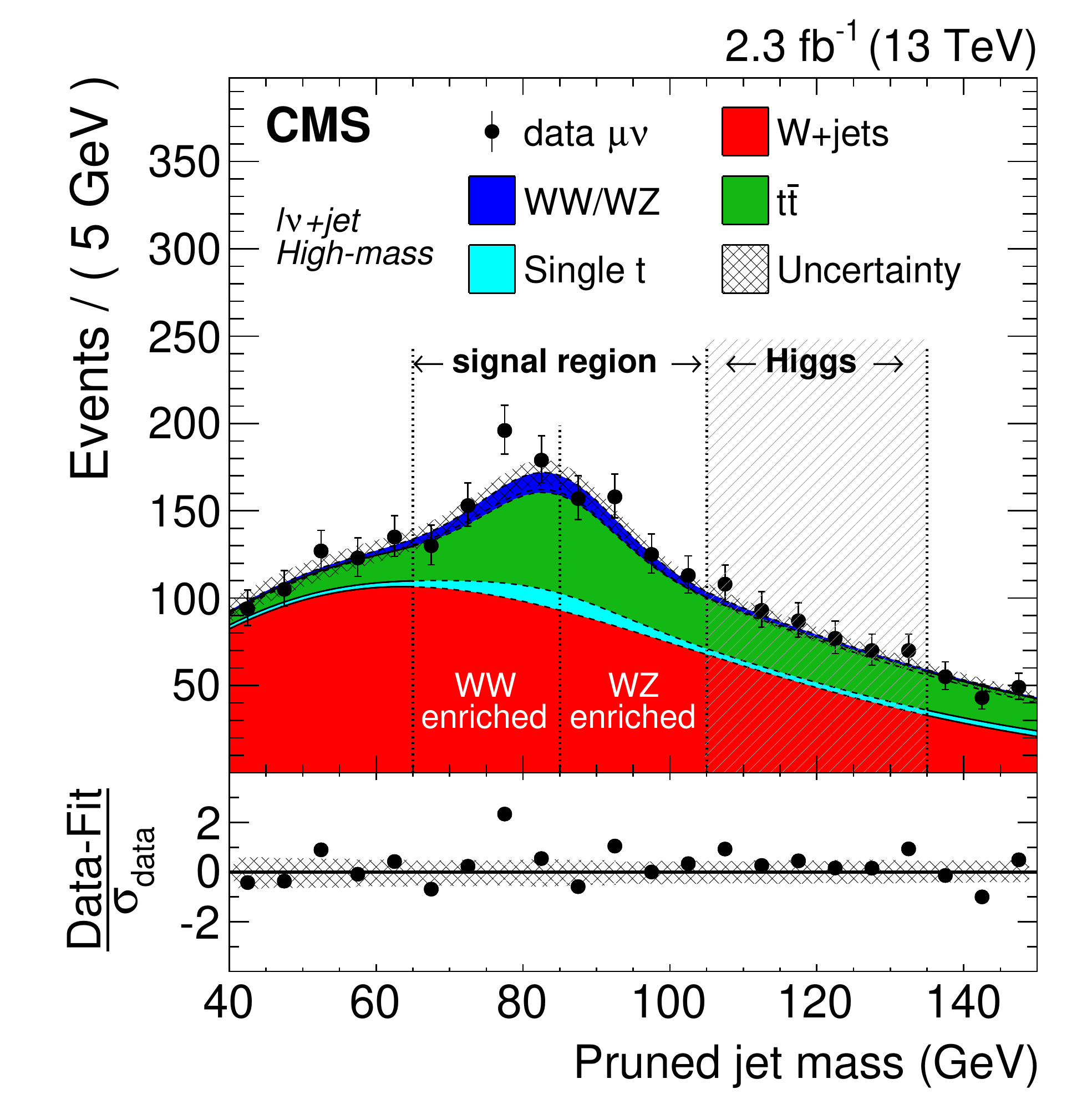}
\includegraphics[width=\cmsFigWidth]{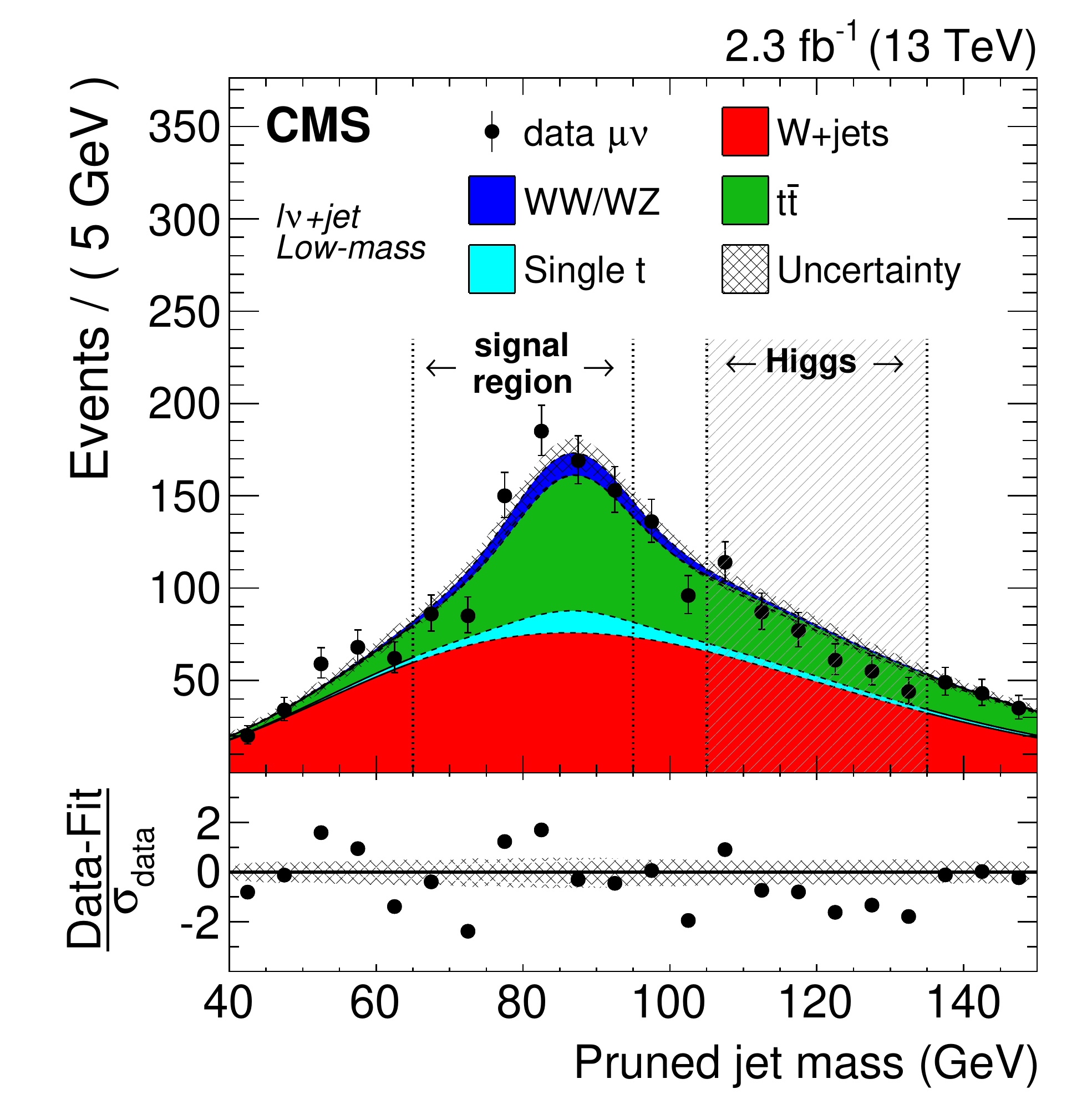}
\caption{
Distributions of the pruned jet mass $\mJ$ in the \lnujet{} high-mass (left) and low-mass (right) analyses in the muon channel.
All selections are applied except the requirement on $\mJ$ signal window.
Data are shown as black points.
The signal regions and $\mJ$ categories of the analyses are indicated by the vertical dotted lines. The shaded $\mJ$ region \SRHIGH--\HIGHSBLOW{\GeV} is not used in these analyses. In the lower panel of each plot, the bin-by-bin fit residuals, $(N_\text{data}- N_\text{fit})/\sigma_\text{data}$, are shown together with the uncertainty band of the fit normalized by the statistical uncertainty of data points, $\sigma_\text{data}$.
}
\label{fig:VJetsNormalization}
\end{figure}

 The overall normalization of the W+jets background in the signal region is determined from a fit to the \mJ{} distribution in the lower and upper sidebands of the data. The analytical form of the fitting function is chosen from simulation studies, as are the contributions from minor backgrounds. Figure~\ref{fig:VJetsNormalization} shows the result of this fit for the low- and high-mass \lnujet{} channels.

The form of the $\mVV$ distribution for the W+jets background in the signal region (SR) is determined from the lower $\mJ$ sideband (SB), through the transfer function $\alpha_\mathrm{MC}(\mVV)$ obtained from the W+jets simulation, and defined as:
\begin{equation}
\alpha_\mathrm{MC}(\mVV) = \frac{F_\mathrm{MC, SR}^{\mathrm{W+jets}}(\mVV)}{F_\mathrm{MC, SB}^{\mathrm{W+jets}}(\mVV)},
\end{equation}
where $F(\mVV)$ is the probability density function used to describe the \mVV spectrum in different regions.
The upper \mJ{} sideband is not considered in this fit since the expected \mVV distribution is different here,
displaying a threshold effect not present in the lower sideband and signal regions.
The adopted parameterization for the $\mVV$ spectrum in both SR and SB regions is of the form
$f(x) \propto \re^{c_0x+c_1/x}$,
which is found to adequately describe the simulation. Tests are performed with alternative functional forms, and the prediction for the backgrounds is found to agree
with the one of the default function within the uncertainties.

The $\mVV$ distribution observed in the lower sideband region is corrected for the presence of minor backgrounds to have an estimate of the W+jets contribution in the control region of the data, \smash{$F_{\text{\tiny DATA}, \text{\tiny SB}}^{\text{\tiny W+jets}}(\mVV)$}. The W+jets background distribution in the signal region is then obtained by rescaling \smash{$F_{\text{\tiny DATA}, \text{\tiny SB}}^{\text{\tiny W+jets}}(\mVV)$} by \smash{$\alpha_\text{\tiny MC}(\mVV)$}. The minor backgrounds are then added to the W+jets background to obtain the total SM prediction in the signal region.

\begin{figure}[htbp]
\centering
\includegraphics[width=\cmsFigWidth]{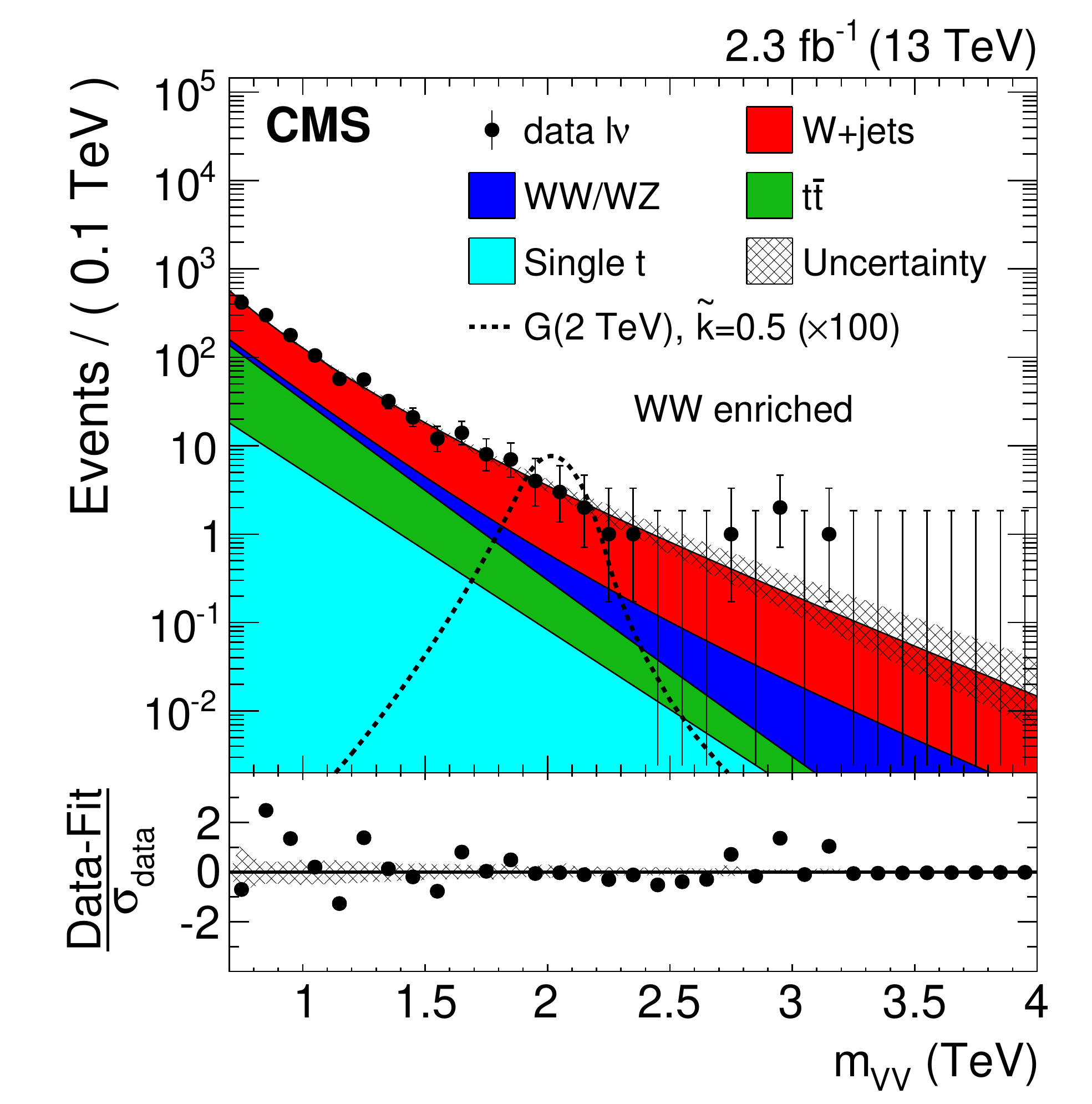}
\includegraphics[width=\cmsFigWidth]{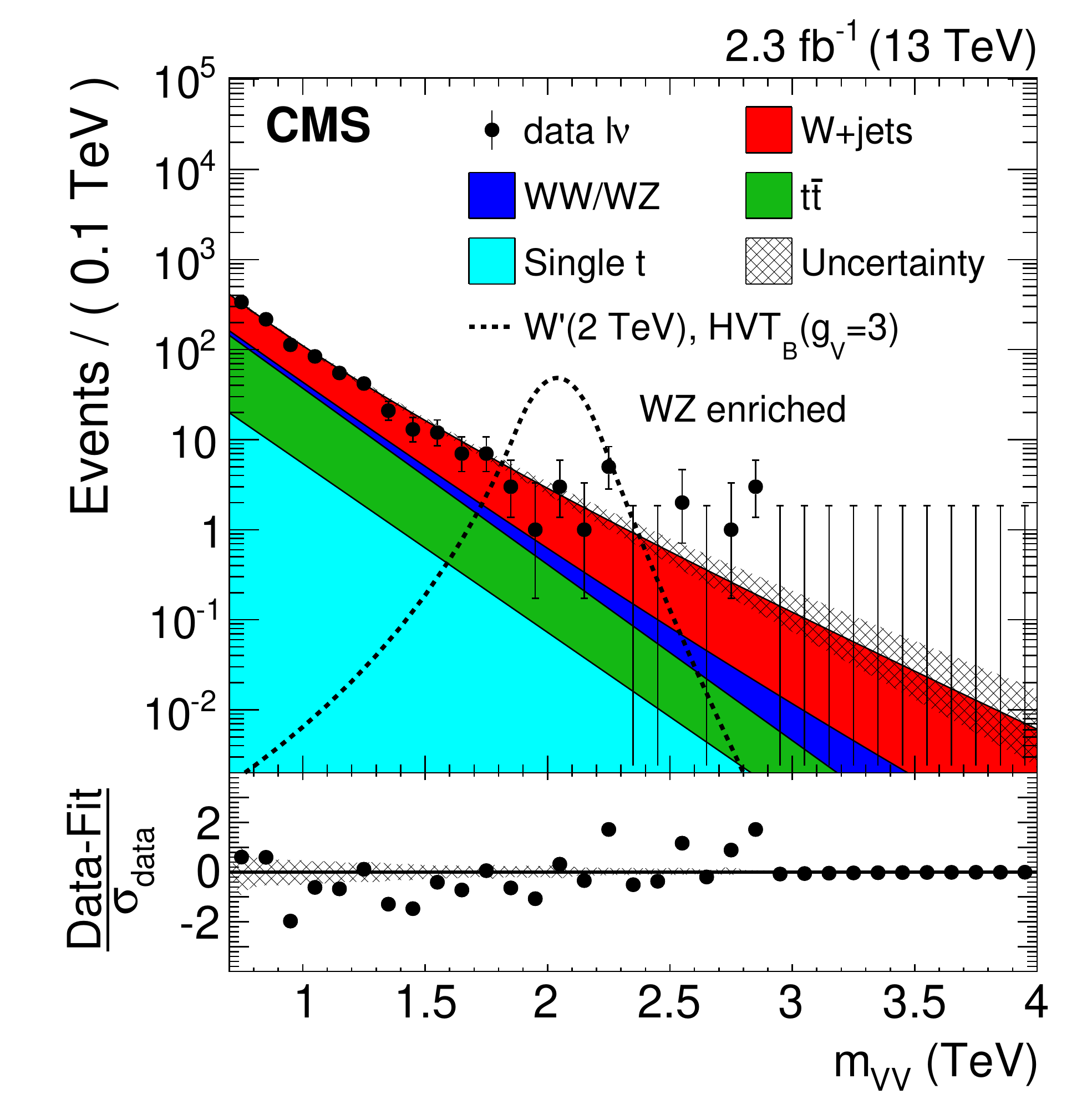}\\
\includegraphics[width=\cmsFigWidth]{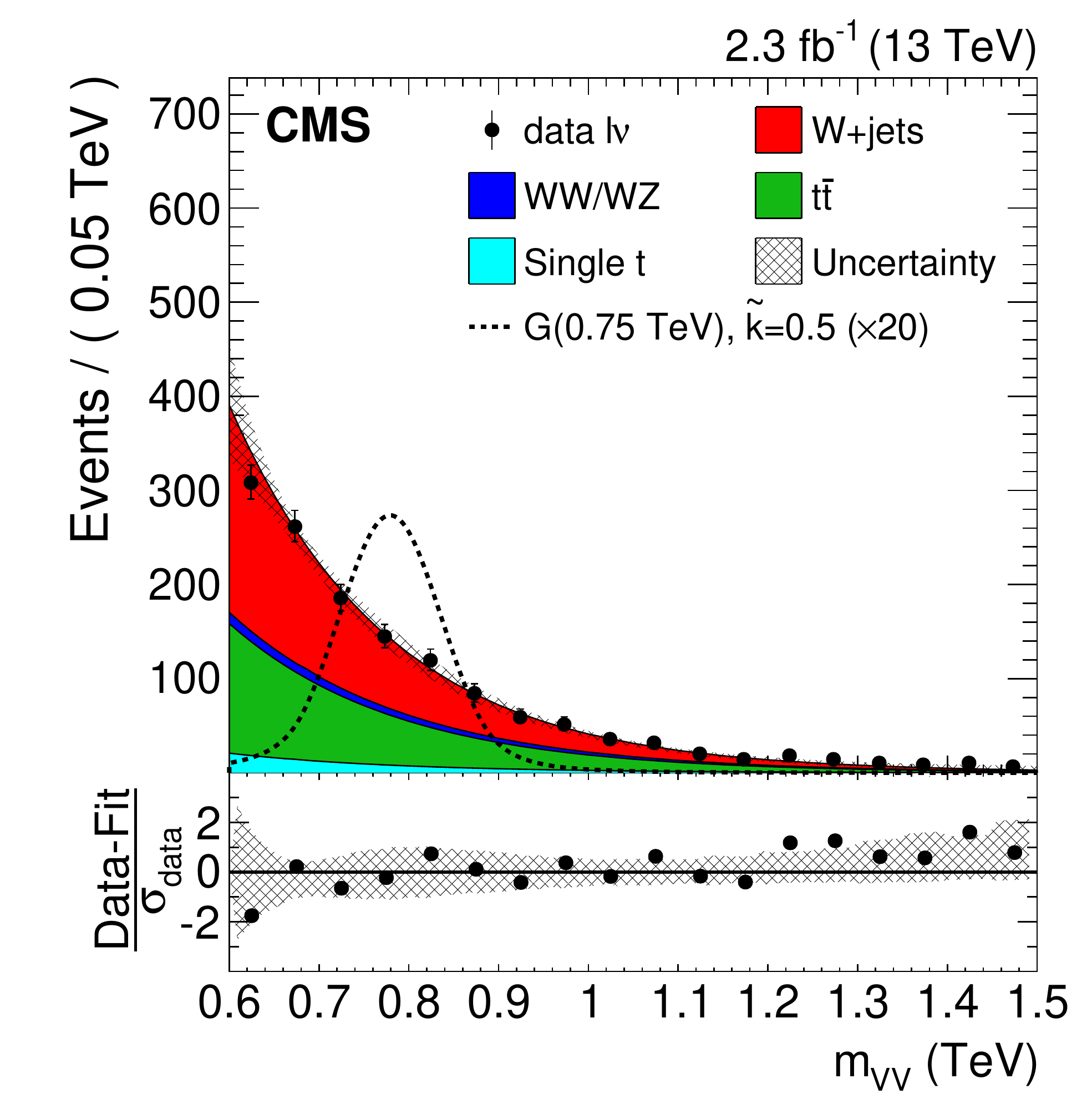}
\caption{
(Upper plots) Final $\mVV$ distributions for data and expected backgrounds in the high-mass analysis obtained from the combined muon and electron channels in the WW-enriched (left) and WZ-enriched (right) signal regions.
(Lower plot) Final $\mVV$ distributions for data and expected backgrounds in the signal region of the low-mass analysis obtained from the combined muon and electron channels.
In each plot the solid curve represents the background estimation provided by the $\alpha$ ratio method.
The hatched band includes both statistical and systematic uncertainties.
The data are shown as black points.
Signal benchmarks for a mass of 2\TeV (0.75\TeV) are also shown with black dashed lines for the upper (lower) plots.
In the lower panel of each plot are the bin-by-bin fit residuals, ($N_\text{data}- N_\text{fit}$)/$\sigma_\text{data}$, shown together with the uncertainty band of the fit normalized by the statistical uncertainty of data, $\sigma_\text{data}$. }
\label{fig:MVV_VW_mu_el}
\end{figure}

Figure~\ref{fig:MVV_VW_mu_el} shows the final spectrum in $\mVV$ for events in all categories for the low- and high-mass analyses. The observed data and the predicted background agree. The highest mass events in the \lnujet channel are at  $\mVV = 2.95$ and 3.15\TeV for the muon and electron categories, respectively.

\subsection{Signal modelling}
\label{sec:signal}

\begin{figure}[!htb]
\centering
\includegraphics[width=0.8\cmsFigWidth]{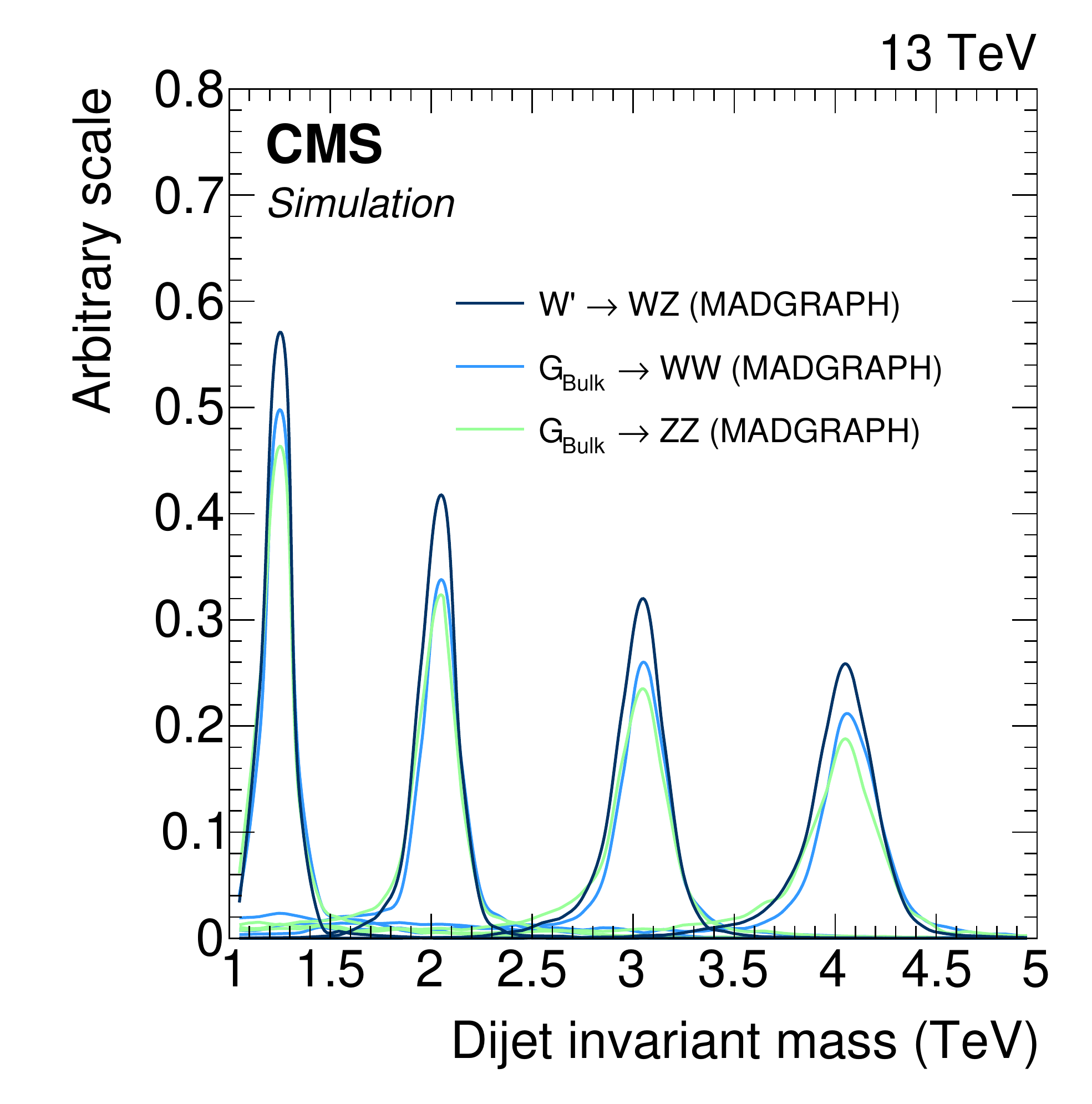}
\includegraphics[width=0.8\cmsFigWidth]{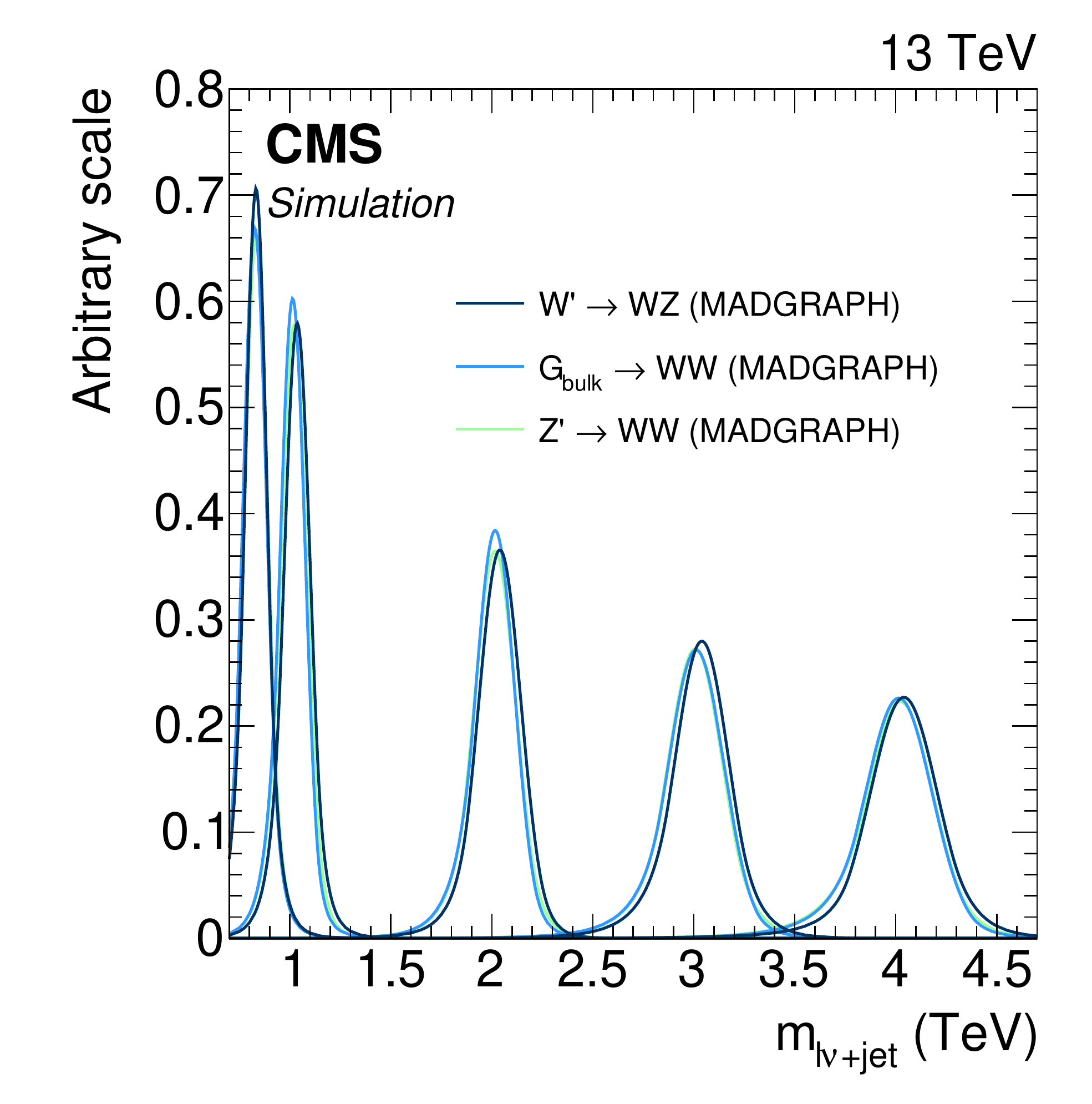}
\caption{Dijet invariant mass (left) and $m_{\ell\nu+{\rm jet}}$ (right) distributions expected for different signal mass hypotheses.}
\label{fig:sigfit}
\end{figure}

Figure~\ref{fig:sigfit} shows the simulated \mjj{} and $m_{\ell\nu+{\rm jet}}$ distributions for different resonance masses from 0.8 to 4.0\TeV.
The experimental resolution for the dijet channel is around 4\%, while it ranges from 6\% at 1\TeV to 4\% at 4\TeV in the \lnujet{} channel.
We adopt an analytical description of the signal, choosing a double-sided Crystal Ball
(CB) function~\cite{CrystalBallRef} (\ie a Gaussian core with power law tails on both
sides) to describe the simulated resonance distributions.
A linear interpolation between a set of reference distributions (corresponding
to masses of 0.6, 0.7, 0.8, 1.0, 1.2, 1.4, 1.6, 1.8, 2.0, 2.5, 3.0, 3.5, and 4.0\TeV) is used
to estimate the expected distributions for intermediate values of resonance mass.
Table~\ref{tab:efficiencies} summarizes the overall event-selection efficiency for our chosen analysis channels and
signal models. All channels are used in the statistical analysis of each signal.

\begin{table}[!htb]
  \centering

  \caption{Summary of signal efficiencies for all analysis channels and all signal models.
The quoted efficiencies are in percent, and include the branching fractions of the two vector bosons to the final state of the analysis channel, effects from detector acceptance, as well as reconstruction and selection efficiencies.
Values are not indicated for categories and masses where the analysis channel has no sensitivity.
}
\label{tab:efficiencies}
  \begin{tabular}{l c c c c c c c c c c c}
    \hline
     & & \multicolumn{10}{c}{Efficiency (\%)} \\
     & & \multicolumn{6}{c}{Dijet channel} & \multicolumn{4}{c}{\lnujet{} channel}\\
     Signal & Mass (TeV) & \multicolumn{2}{c}{WW} & \multicolumn{2}{c}{WZ} & \multicolumn{2}{c}{ZZ} & \multicolumn{2}{c}{WW} & \multicolumn{2}{c}{WZ}\\
     & & HP & LP & HP & LP & HP & LP & e & $\mu$ & e & $\mu$ \\
    \hline
$\BulkG \to \PW\PW$ & 0.75 & \NA   & \NA   & \NA   & \NA   & \NA   & \NA   & 4.4  & 5.3  & \NA   & \NA \\
$\BulkG \to \PW\PW$ & 1.2  & 4.9 & 5.6 & 3.6 & 3.9 & 0.6 & 0.6 & 5.7  & 7.4  & 1.7 & 2.1\\
$\BulkG \to \PW\PW$ & 2.0  & 6.5 & 9.1 & 2.1 & 2.9 & 0.2 & 0.3 & 7.3  & 8.0  & 1.4 & 1.5\\
$\BulkG \to \PW\PW$ & 3.0  & 4.9 & 7.8 & 2.3 & 3.3 & 0.3 & 0.3 & 7.0  & 7.5  & 1.5 & 1.7\\
$\BulkG \to \PW\PW$ & 4.0  & 4.2 & 8.0 & 2.8 & 3.9 & 0.3 & 0.6 & 7.0  & 7.0  & 2.0 & 1.9\\
    \hline
$\BulkG \to \Zo\Zo$ & 1.2 & 1.1 & 1.2 & 5.3 & 5.1 & 6.1 & 4.6 & \NA & \NA & \NA & \NA\\
$\BulkG \to \Zo\Zo$ & 2.0 & 1.3 & 2.3 & 5.0 & 6.7 & 4.7 & 4.5 & \NA & \NA & \NA & \NA\\
$\BulkG \to \Zo\Zo$ & 3.0 & 1.1 & 2.5 & 4.3 & 7.2 & 3.8 & 4.5 & \NA & \NA & \NA & \NA\\
$\BulkG \to \Zo\Zo$ & 4.0 & 0.9 & 2.7 & 3.7 & 7.2 & 3.7 & 4.3 & \NA & \NA & \NA & \NA\\
    \hline
HVT $\PWpr \to \PW\Zo$ & 0.75 & \NA   & \NA   & \NA   & \NA   & \NA   & \NA   & 1.3 & 1.6 & \NA   & \NA \\
HVT $\PWpr \to \PW\Zo$ & 1.2  & 2.7 & 3.0 & 7.2 & 6.8 & 1.5 & 1.4 & 1.2 & 1.6 & 2.8 & 3.4\\
HVT $\PWpr \to \PW\Zo$ & 2.0  & 3.0 & 4.7 & 6.0 & 6.7 & 0.8 & 0.8 & 1.8 & 2.0 & 3.0 & 3.3\\
HVT $\PWpr \to \PW\Zo$ & 3.0  & 2.3 & 4.5 & 5.0 & 6.8 & 1.0 & 0.8 & 1.9 & 2.0 & 3.1 & 3.2\\
HVT $\PWpr \to \PW\Zo$ & 4.0  & 1.9 & 4.2 & 4.2 & 6.4 & 1.0 & 1.2 & 1.9 & 2.0 & 3.1 & 3.0\\
    \hline
HVT $\PZpr \to \PW\Wo$ & 0.75 & \NA   & \NA   & \NA   & \NA   & \NA   & \NA   & 4.1  & 5.1  & \NA   & \NA \\
HVT $\PZpr \to \PW\Wo$ & 1.2  & 7.2 & 7.6 & 3.3 & 3.6 & 0.4 & 0.4 & 6.0  & 7.7  & 1.6 & 2.0\\
HVT $\PZpr \to \PW\Wo$ & 2.0  & 6.1 & 8.1 & 2.0 & 2.3 & 0.1 & 0.2 & 7.9  & 8.8  & 1.3 & 1.5\\
HVT $\PZpr \to \PW\Wo$ & 3.0  & 4.7 & 8.0 & 2.1 & 2.8 & 0.3 & 0.2 & 7.5  & 8.1  & 1.6 & 1.5\\
HVT $\PZpr \to \PW\Wo$ & 4.0  & 3.8 & 6.7 & 2.1 & 3.0 & 0.3 & 0.3 & 7.4  & 7.6  & 1.9 & 1.9\\
    \hline
  \end{tabular}
\end{table}

\section{Systematic uncertainties}
\label{sec:systematicuncertainties}

\subsection{Systematic uncertainties in the background estimation}

\par For the dijet analysis, the background estimation is obtained from a fit to the data. As such, the only relevant uncertainty is the statistical one as represented by the covariance matrix of the fit to the dijet function. Different parameterizations of the fitting function have been studied, and the differences observed are well within the bounds of the aforementioned uncertainty and are  assumed to pose no additional contribution.

\par For the \lnujet{} analyses, uncertainties in both the distribution and normalization of the background prediction can be important. The uncertainty in the distribution is dominated by the statistical uncertainties in the simultaneous fits to the data of the sideband region, and the simulation in signal and sideband regions. An effect of almost equal magnitude is due to the uncertainties in the modelling of the transfer function $\alpha(\mVV)$ between the sideband and the signal region. The uncertainty in the normalization of the background has three sources: the W+jets component, dominated by the statistical uncertainty of the events in the pruned jet mass sideband, varying from 5 to 9\%; the \ttbar/single top quark component, dominated by the scale factor obtained from the top quark enriched control region, amounting to about 5--7\% and 8\% in the muon and electron channels, respectively; and the diboson component, dominated by the V tagging uncertainty, which varies in the range of 3--25\%.

\subsection{Systematic uncertainties in the signal prediction}

\par The dominant uncertainty in the signal selection efficiency arises from uncertainties in data-to-simulation scale factors for the V tagging efficiency derived from a top quark enriched control sample, as described in Section~\ref{subsec:Vhadr}. The normalization uncertainties are summarized in Tables~\ref{tab:VV_systematicssummary_signal} and~\ref{tab:VW_systematicssummary_signal} for the dijet and \lnujet{} channels, respectively.

\par Uncertainties in the reconstruction of jets affect both the signal efficiency and the distribution in the reconstructed resonance mass. The four-momenta of the reconstructed jets are rescaled or smeared according to the uncertainties in the respective jet energy scale or resolution. The selection efficiencies are recalculated on these modified events, with the resulting changes taken as systematic uncertainties that depend on the resonance mass. The induced changes on the reconstructed resonances are propagated as uncertainties in the peak position and width of the Gaussian core.
In addition, the induced relative migration among V jet mass categories is evaluated, and found not to affect the overall signal efficiency.
The correlations in these uncertainties between the different categories are taken into account.

\par The uncertainty in the lepton energy scale is correlated with the obtained signal efficiency.
Changes in lepton energy are propagated to the reconstructed \PTm, and through the entire analysis. The relative change in the number of selected signal events is taken as a systematic uncertainty in the signal normalization. For both lepton flavors, the uncertainties are smaller than 1\%,
and are uncorrelated for different lepton flavors, but correlated for different pruned jet mass and $\nsubj$ categories.
In addition, the induced change in peak position and its width are added as systematic uncertainties in the distribution of the signal.
Again, for both lepton flavors, the uncertainties are below 1\%.

\par The systematic uncertainties in the lepton trigger, identification, and isolation efficiencies are obtained using a tag-and-probe method in $\PZ \to \ell\ell$ events~\cite{CMS:FirstInclZ}, and are used only for the \lnujet{} channel. An uncertainty of 1--3\% is assigned to the trigger efficiency for both lepton flavors, depending on the lepton \pt and $\eta$. For lepton identification and isolation efficiency, the systematic uncertainty is estimated to be 1--2\% for the muon and 3\% for electron flavors.

\par The \LUMIUNCERT uncertainty in the integrated luminosity~\cite{CMS-PAS-LUM-15-001} applies to the normalization of signal events.
Uncertainties in the signal yield due to the choice of PDFs and the values chosen for the factorization ($\mu_{f}$) and renormalization ($\mu_{r}$) scales are also taken into account.
The PDF uncertainties are evaluated using the NNPDF 3.0~\cite{Ball:2011mu} PDFs.
The uncertainty related to the choice of $\mu_{f}$ and $\mu_{r}$ scales is evaluated following the proposal in Refs.~\cite{Cacciari:2003fi,Catani:2003zt} by varying the default choice of scales in the following 6 combinations of factors:
$(\mu_{f}$, $\mu_{r})$ $\times$ $(1/2, 1/2)$, $(1/2, 1)$, $(1,1/2)$, $(2, 2)$, $(2, 1)$, and $(1, 2)$.
The uncertainty in the signal cross section from the choice of PDFs and of factorization and renormalization scales ranges from 4 to 77\%, and from 1 to 22\%, respectively, depending on the resonance mass, particle type and its production mechanism.
These uncertainties are fully correlated among the \lnujet{} and dijet channels.

Tables~\ref{tab:VV_systematicssummary_signal} and \ref{tab:VW_systematicssummary_signal} summarize the systematic uncertainties in the dijet and \lnujet{} channels, respectively.

\begin{table}[htb]
\topcaption{Summary of the systematic uncertainties in the contribution from signal in the dijet analysis and their impact on the event yield in the signal region and on the
 reconstructed distribution in \mVV{} (mean and width). The last three uncertainties result in migrations between event categories, but do not affect the overall signal efficiency.}
  \centering
  \begin{tabular}{lccc}
    \hline
    Source \T                        & Relevant quantity      & HP uncertainty (\%)  & LP uncertainty (\%)\\
    \hline
    Jet energy scale                 & Resonance shape        & 2                    & 2 \\
    Jet energy resolution            & Resonance shape        & 10                   & 10 \\
    \hline
    Jet energy and \mJ{} scale       & Signal yield           & \multicolumn{2}{c}{0.1--4}\\
    Jet energy and \mJ{} resolution  & Signal yield           & \multicolumn{2}{c}{0.1--1.4}\\
    Pileup                           & Signal yield           & \multicolumn{2}{c}{2}\\
    Integrated luminosity            & Signal yield           & \multicolumn{2}{c}{2}\\
    PDFs (\PWpr)                     & Signal yield		      & \multicolumn{2}{c}{4--19}\\
    PDFs (\PZpr)                     & Signal yield		      & \multicolumn{2}{c}{4--13}\\
    PDFs (\BulkG)                    & Signal yield		      & \multicolumn{2}{c}{9--77}\\
    Scales (\PWpr)                   & Signal yield		      & \multicolumn{2}{c}{1--14}\\
    Scales (\PZpr)                   & Signal yield		      & \multicolumn{2}{c}{1--13}\\
    Scales (\BulkG)                  & Signal yield		      & \multicolumn{2}{c}{8--22}\\
    \hline
    Jet energy and \mJ{} scale       & Migration              & \multicolumn{2}{c}{1--50}\\
    V tagging \nsubj{}               & Migration              & 14                    & 21\\
    V tagging \pt-dependence         & Migration              & 7--14                & 5--11\\
    \hline
  \end{tabular}
  \label{tab:VV_systematicssummary_signal}
\end{table}

\begin{table}[htb]
\topcaption{Summary of the signal systematic uncertainties for the \lnujet{} analyses and their impact on the event yield in the signal region and on the
 reconstructed \mVV{} shape (mean and width) for both muon and electron channels.
The last three uncertainties result in migrations between event categories, but do not affect the overall signal efficiency.
The correlations among different categories are taken into account.}
\centering
  \begin{tabular}{lccc}
    \hline
    Source \T                           & Relevant quantity          & Uncertainty (\%)\\
	\hline
	Lepton trigger ($\mu$/e) 	        & Signal yield		         & 1--3 / 1--3\\
	Lepton identification	($\mu$/e)	& Signal yield		         & 1--2 / 3\\
	Jet energy and \mJ{} scale          & Signal yield		         & 0.2--4 \\
	Jet energy and \mJ{} resolution     & Signal yield		         & 0.1--2 \\
    Integrated luminosity		        & Signal yield		         & 2.7\\
    PDFs (\PWpr)                     & Signal yield		      & 4--19\\
    PDFs (\PZpr)                     & Signal yield		      & 4--13\\
    PDFs (\BulkG)                    & Signal yield		      & 9--77\\
    Scales (\PWpr)                   & Signal yield		      & 1--14\\
    Scales (\PZpr)                   & Signal yield		      & 1--13\\
    Scales (\BulkG)                  & Signal yield		      & 8--22\\
	\hline
	Jet energy scale		            & Resonance shape (mean)	 & 1.3\\
	Jet energy scale		            & Resonance shape (width)	 & 2--3\\
	Jet energy resolution	            & Resonance shape (mean)	 & 0.1\\
	Jet energy resolution		        & Resonance shape (width)	 & 4\\
    \hline
    Jet energy and \mJ{} scale          & Migration                  & 2--24\\
	V tagging \nsubj{} (0.45/0.6)       & Migration 	             & 7 / 3\\
    V tagging \pt-dependence (0.45/0.6) & Migration                  & 3--6 / 6--10\\
	\hline
    \end{tabular}
\label{tab:VW_systematicssummary_signal}
\end{table}

\section{Statistical interpretation}
\label{sec:statisticalinterpretation}

The \mVV{} distribution observed in data and the
SM background prediction are compared to check for the presence of a
new resonance decaying to vector bosons.
No bins with an excess with significance larger than three standard deviations are observed.
We set upper limits on the
production cross section of such resonances by combining the event categories of
the dijet and \lnujet analyses.
We follow the asymptotic approximation~\cite{AsymptCLs} of the $\mathrm{CL_S}$ criterion described in Refs.~\cite{CLs1,Junk:1999kv}.
The limits computed following this approach are found to agree with the results obtained using the modified frequentist prescription~\cite{CLs1,Junk:1999kv}.
For each channel and each signal hypothesis a likelihood function is built from the reconstructed \mVV mass distribution observed in data, the background prediction, and the
signal resonance shape. A maximum-likelihood fit to the data is then performed to obtain the best estimate of the signal cross section.
Systematic uncertainties are profiled~\cite{Aad:2015zhl} as log-normal nuisance parameters in the statistical interpretation,
and for each possible value of the fitted signal cross section they are all refitted to maximize the likelihood.

\subsection{ Limits on narrow-width resonance models}
\label{subsec:narrow-results}

Exclusion limits are set in the context of the bulk graviton model and of the HVT Models A and B,
under the assumption of a natural width negligible compared to the
experimental resolution (narrow-width approximation).
To maximize the sensitivity of the search we combine the results from all the analysis channels
in each of the considered signal hypotheses. In the combination, the systematic uncertainties in jet momentum scale and resolution, V tagging efficiency scale factors, and integrated luminosity are assumed to be 100\% correlated.

\begin{figure}[!htb]
\centering
     \includegraphics[width=\cmsFigWidth]{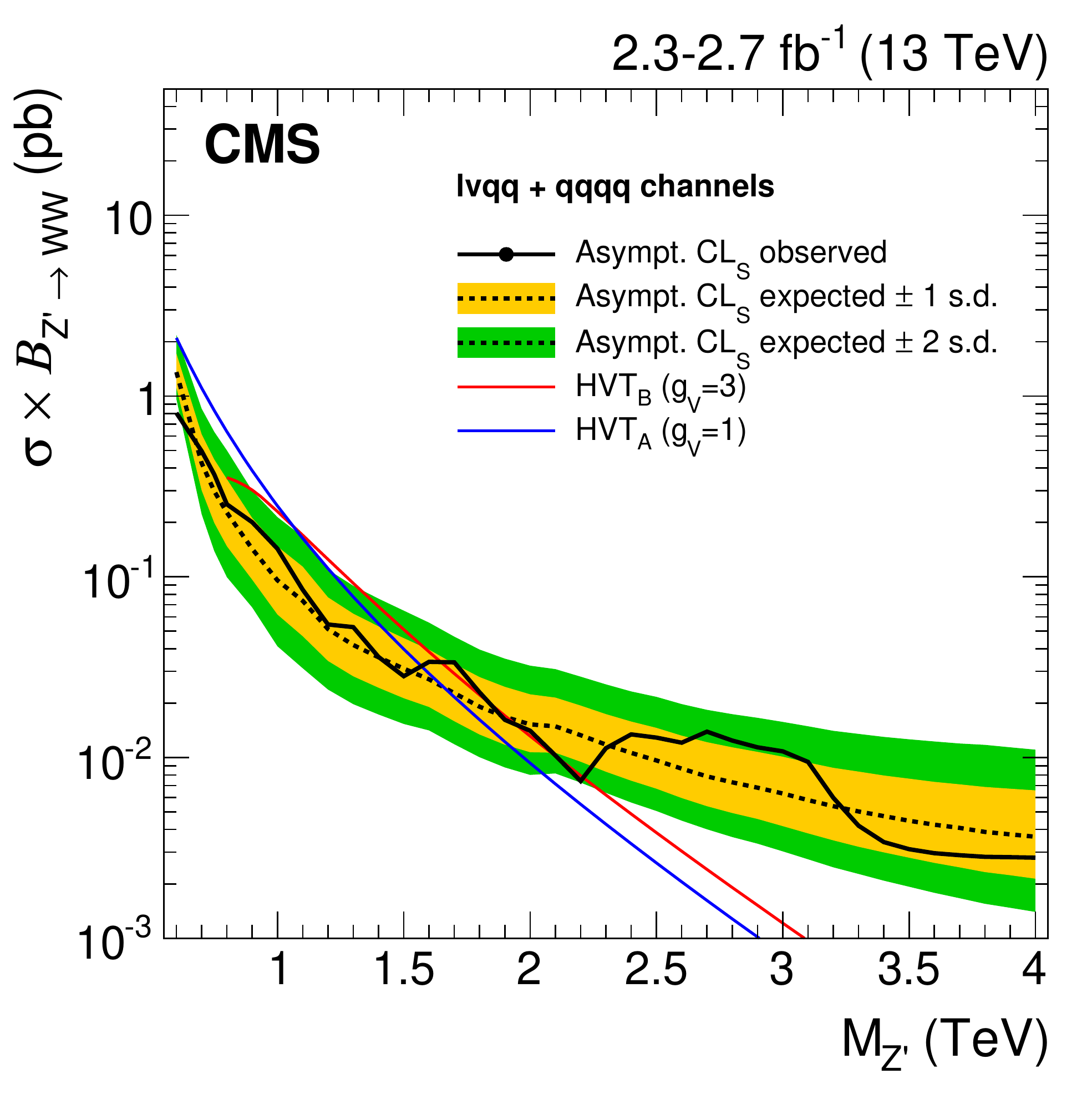}
     \includegraphics[width=\cmsFigWidth]{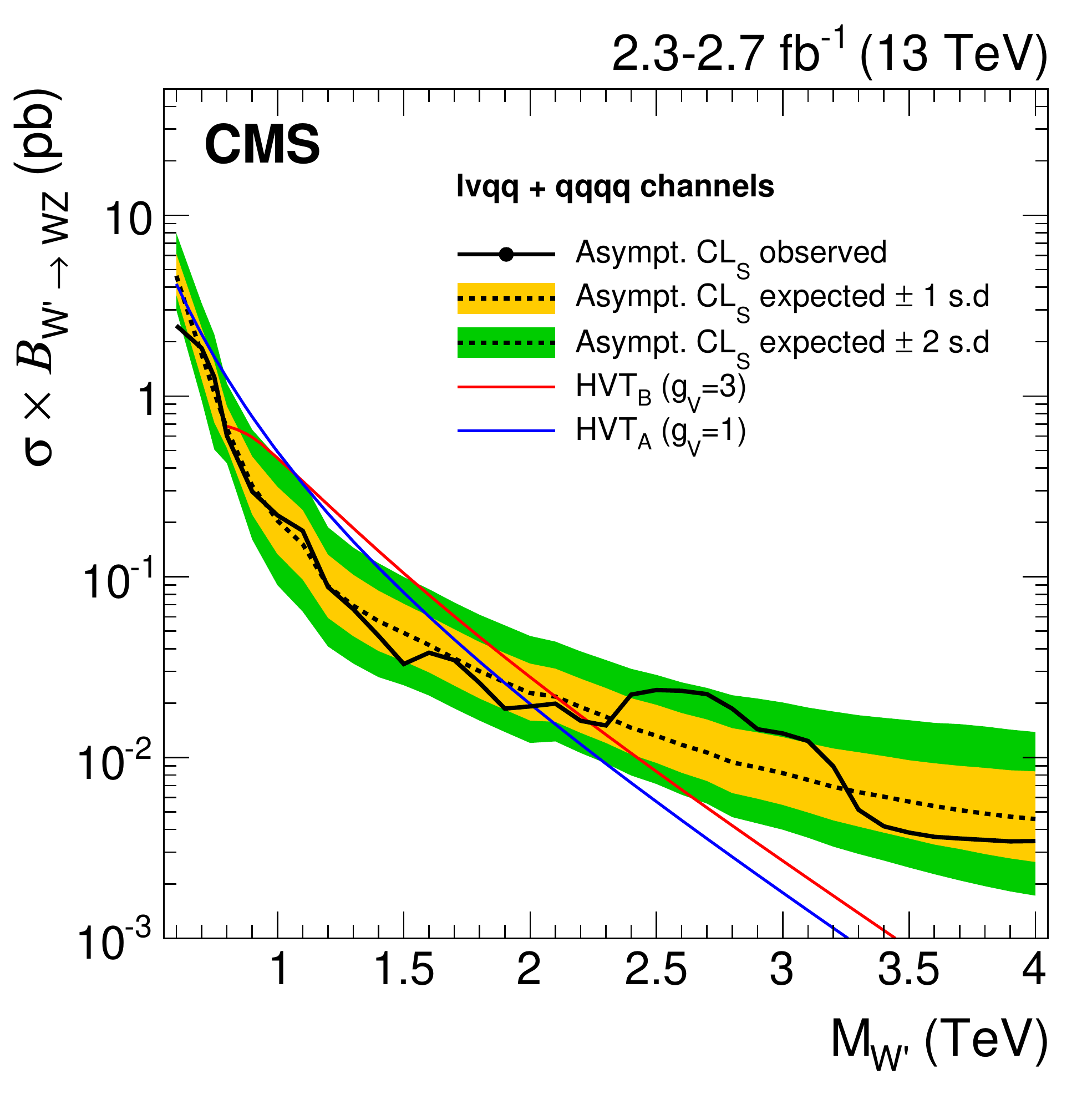}\\
     \includegraphics[width=\cmsFigWidth]{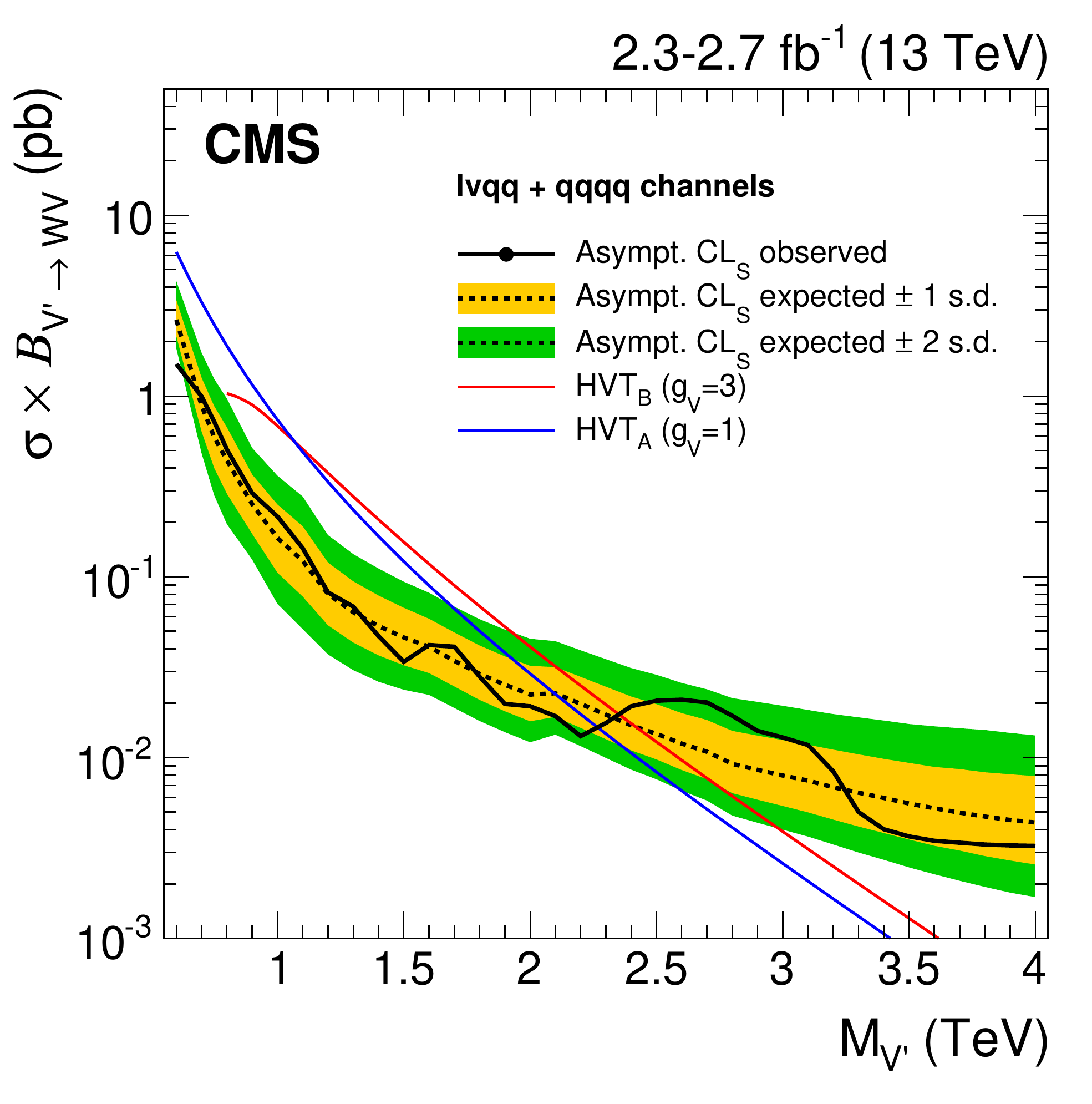}
     \includegraphics[width=\cmsFigWidth]{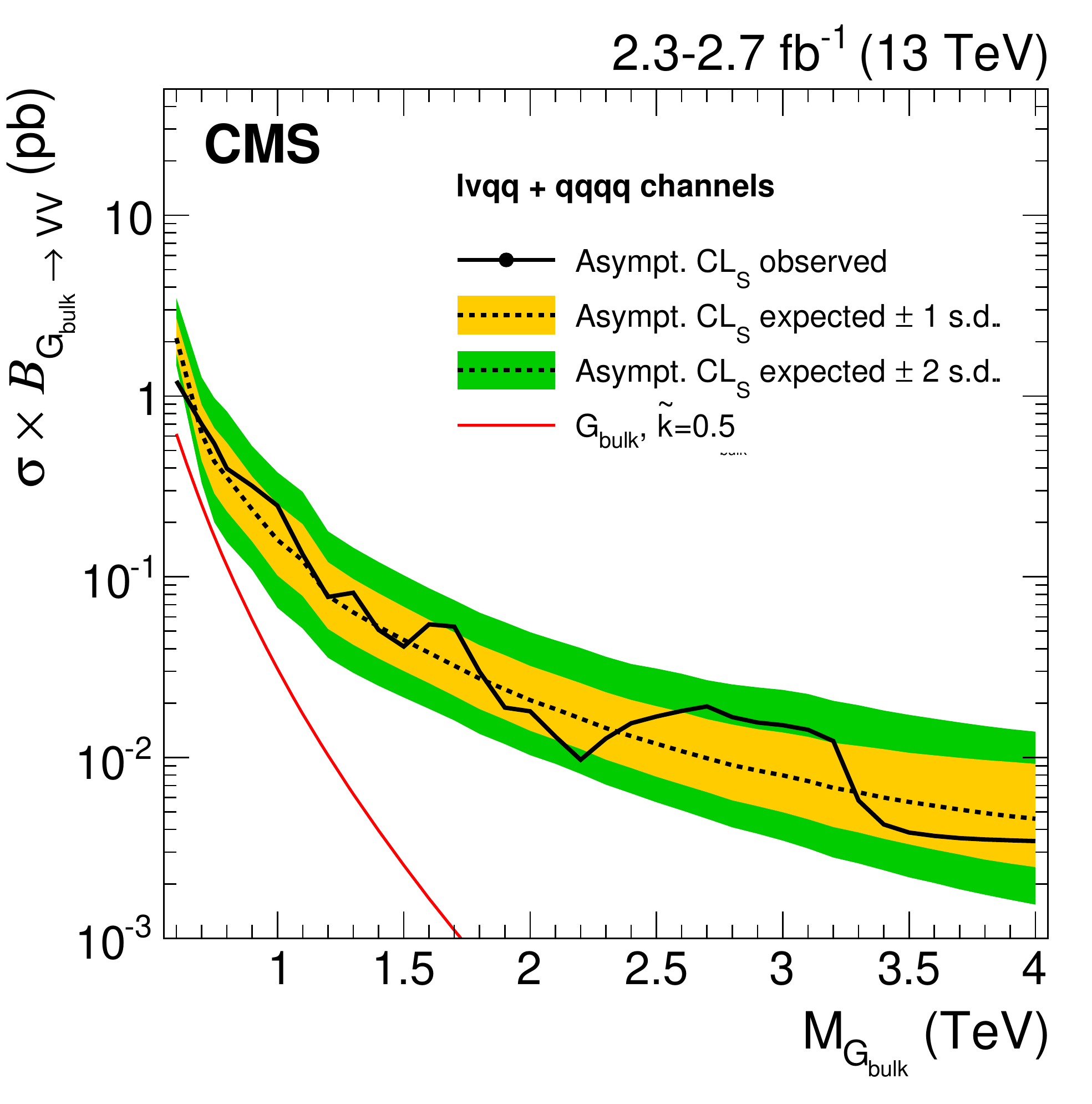}
\caption{Observed (black solid) and expected (black dashed) 95\% CL upper limits on the production of a narrow-width resonance decaying to
a pair of vector bosons for different signal hypotheses. In the upper plots, limits are set in the context of a spin-1 neutral \PZpr (left) and charged \PWpr (right)
resonances, and compared with the prediction of the HVT Models A and B. In the lower left plot, limits are set in the same model under the triplet hypothesis (\PWpr and \PZpr).
In the lower right plot, limits are set in the context of a bulk graviton with $\ktilde =0.5$ and compared with the prediction.
For $\mathrm{G}_\text{bulk}$, \PZpr and triplet signals (W' signal) with masses $<$0.8\TeV ($<$0.75\TeV), the limits are obtained from the low-mass \lnujet channel, while for the higher masses they are obtained from the high-mass \lnujet and dijet channels.
}
\label{fig:limitCombined}
\end{figure}

Figure~\ref{fig:limitCombined} shows the resulting expected and observed
exclusion limits at 95\% CL on the signal cross section as a function of the resonance mass for all signal hypotheses.
The limits are compared with the product of cross section and branching fraction ($\sigma\mathcal{B}$) to $\Wo\Wo$ and $\Zo\Zo$
for a bulk graviton with $\ktilde = 0.5$, and with $\sigma\mathcal{B}$ for $\Wo\Zo$ and $\Wo\Wo$ for spin-1 particles predicted by the HVT Models A and B.
In this context, we consider a scenario where we expect the \PWpr and \PZpr bosons to be degenerate in mass (triplet hypothesis).
In addition, we consider also a scenario where only a charged (\PWpr) or a neutral (\PZpr) resonance is expected at a given mass (singlet hypothesis).
Combining the analyses leads to about 10--30\% more stringent expected upper limits on the cross section compared
to the most sensitive individual channel, depending on the resonance mass and the signal hypothesis.
For $\mathrm{G}_\text{bulk}$, \PZpr and triplet signals (W' signal) with masses $<$0.8\TeV ($<$0.75\TeV), the limits are obtained from the low-mass \lnujet channel, while for the higher masses they are obtained from the high-mass \lnujet and dijet channels.
The dijet analysis provides more stringent expected upper limits on the cross sections than the \lnujet{} analysis for resonance masses above 1.7\TeV for \PZpr and $>$1.3\TeV for \PWpr, because of the larger branching fractions: $\mathcal{B}(\Wo\Wo\to\qqbar\qqbar) = 44\%$, $\mathcal{B}(\Wo\Wo\to\ell\Pgn\qqbar) = 31\%$, $\mathcal{B}(\Wo\Zo\to\qqbar\qqbar) = 46\%$, and $\mathcal{B}$($\Wo\Zo\to\ell\Pgn\qqbar$) = 16\%. In fact, the combination of high- and low-purity categories, together with the weak dependence of tagging efficiency on \pt~\cite{CMS-PAS-JME-14-002} improves the sensitivity for most potential signal models.
In the narrow-width bulk graviton model, the combined sensitivity of the searches is
not large enough to set mass limits, however, cross sections are excluded in the range 3--1200\unit{fb}.
For HVT Model A (B), the combined data exclude singlet \PWpr resonances with masses $<$2.0~(2.2)\TeV and \PZpr resonances with masses below $<$1.6~(1.7)\TeV.
Under the triplet hypothesis, spin-1 resonances with masses $<$2.3 and $<$2.4\TeV are excluded for HVT Models A and B, respectively.

\begin{figure}[!htb]
\centering
     \includegraphics[width=\cmsFigWidth]{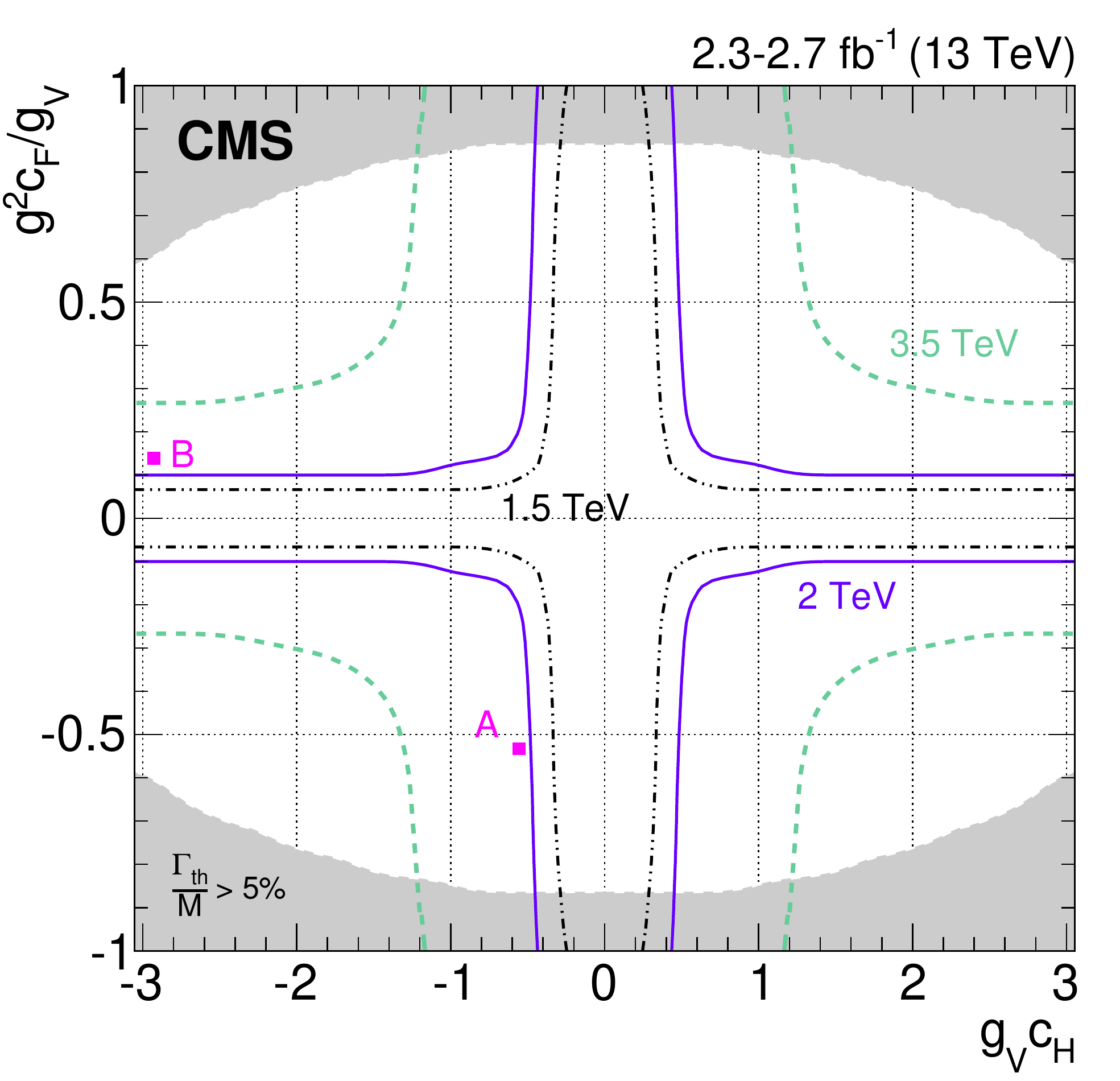}
\caption{
Exclusion regions in the plane of the HVT couplings ($g^2c_\mathrm{F}/g_{\PV},g_{\PV}c_\mathrm{H}$) for three
resonance masses, 1.5, 2.0, and 3.5\TeV. Model points A and B of the benchmarks used in the analysis are also shown.
The solid, dashed, and dashed-dotted lines represent the boundaries of the regions excluded by this search for different resonance masses (the region outside these lines is excluded).
The areas indicated by the solid shading correspond to
regions where the resonance width is predicted to be more than 5\% of the resonance mass and
the narrow-resonance assumption is not satisfied.}
\label{fig:hvtscan}
\end{figure}

Figure~\ref{fig:hvtscan} shows a scan of the coupling parameters and the corresponding observed 95\% CL exclusion
contours in the HVT model for the combined analyses. The parameters
are defined as $g_{\PV}c_\mathrm{H}$ and $g^2c_\mathrm{F}/g_{\PV}$,  related to the coupling strengths of the new resonance to
the Higgs boson and to fermions. The range of the scan is limited by the assumption that the
new resonance is narrow.  A contour is overlaid, representing the region where the theoretical
width is larger than the experimental resolution of the searches, and hence where the narrow-resonance assumption is not satisfied.
This contour is defined by a predicted resonance width of 5\%, corresponding to the narrowest resonance mass resolution of the searches.

\subsection{Model-independent limits}
\label{subsec:mod-indep-results}

The above analysis is specific to a narrow
bulk graviton and HVT models, but these are not the only extension of the SM that predicts resonances decaying to vector bosons. It is therefore useful to reinterpret
these results through a more generic model. In this section, we present the
exclusion limits on the number of events that remain after modifying the analysis and greatly
simplifying its structure, at a moderate cost in
performance. Together with the upper limits on the number of signal
events, we provide tables on reconstruction and identification
efficiencies for vector bosons emitted in the kinematic acceptance of the
analysis. Following the instructions detailed in Appendix~\ref{sec:mod-indep-instr}, it is
possible to estimate the number of
events expected in a generic signal that would be
detected in CMS with the present data set, and to compare it with the
upper limit on the number of signal events.

To avoid the dependence on assumptions in the construction of the
separate categories, we perform a simplified analysis, reducing
the event classification to two (\lnujet{}) and one (dijet) categories, respectively.
This is done by eliminating the low-purity categories and combining the jet mass categories in the analyses. The loss in performance is very
small for a large range of masses.  The effect of dropping the LP category is observed
only at very high masses, where the upper limit on the cross section becomes less stringent.

A generic model cannot be restricted to narrow signals, and we therefore provide limits as a function of both mass (\mX) and natural width (\wX)
of the new resonance. The generated line shape is parametrized with a BW function and its full width at half maximum is defined as the
$\Gamma$ parameter of the BW function. The BW line shape is convolved with a double sided CB function
describing the
detector resolution in the \lnujet{} analysis, and with a sum of a Gaussian and CB functions for the dijet analysis.
As \wX{} is varied, the
parameters of the double-CB function are kept fixed to the values determined under
the narrow-width approximation. It has been checked that the
parametrization of detector effects factorizes from the natural
width of the resonance and is stable as \wX{} increases. The width is scanned
at regular steps of the relative width, $\wX / \mX$, which
spans from the zero-width approximation
(as in the nominal analysis), up to $\wX/\mX=0.30$, in steps of 0.05.
For high masses, the resonance shape is distorted from the BW shape owing to PDF effects creating a tail towards low masses.
The line shape is corrected for this by a linear function that works well for quark induced processes.
However, the shape description using this approach is unsatisfactory for gluon induced processes at very high masses and widths.

We provide the efficiency as a function of the kinematic variables of the vector boson,
as the efficiency can depend significantly on the production
and decay kinematic quantities of the new resonance. The efficiencies are extracted from
the bulk graviton samples generated for the baseline analysis. The
efficiencies are calculated by first preselecting simulated signal
events according to the acceptance requirements of the analysis. The
tables are therefore valid only within this kinematic region, as
summarized in Tables~\ref{tab:GeneratorSelectionsSemileptonic} and~\ref{tab:AllHadGeneratorSelections}
of Appendix~\ref{sec:mod-indep-instr} for the \lnujet{} and dijet analyses,
respectively. For preselected events, the reconstructed V candidates
are then required to pass all the analysis selections.
The efficiencies are presented as a function of the \pt and $\eta$
of the V boson prior to any simulation of detector effects.
All the reweighting and rescaling effects (including
lepton identification and trigger efficiencies, and V tagging scale
factors) are included in the efficiencies.

The efficiencies of requiring no additional well-identified leptons and b-tagged jets
in the \lnujet analysis are found to be independent of the diboson event kinematics.
We use a constant efficiency of 95\% for the combined vetoes. Similarly, the $\Delta\eta$ requirement in the dijet analysis is taken into account as a global efficiency factor of 98\%.

It has been checked that the dependence of the
total signal efficiency and acceptance on the width of
the generated sample is very weak. We
include this effect in the systematic uncertainties of the procedure, as discussed below.

Special consideration is given to cases where the boson is transversely
polarized, because the calculated efficiencies are based on
longitudinally polarized bosons, as in the case of the reference
bulk graviton model. The efficiency of the V tagging selections depend
significantly on the degree of polarization of the vector
boson~\cite{Khachatryan:2014vla}.  This effect is investigated using RS1
gravitons produced with the \MADGRAPH generator. The V bosons
originating from the decays of RS1 gravitons are transversely polarized in
about 90\% of the cases. For bosons decaying leptonically, the
tables are still valid because of the generator-level
selection on individual leptons, which guarantees that
polarization effects for the leptonic boson are included in the
acceptance. As shown in Ref.~\cite{Khachatryan:2014vla}, the efficiency of the jet
substructure selection is found to be smaller for transversely
polarized V bosons that tend to have more asymmetric subjet \pt, resulting in a higher probability for the subjet with lower \pt to be rejected by the pruning algorithm.
Studies of simulated RS1 graviton samples show
that the loss in efficiency is largely independent of the V
kinematic variables, so that the effect of the transverse polarization can be
adequately modelled by a constant scale factor of 0.76, independent of the \pt and $\eta$ of the $\Vo\to\qqbar$ decays.

\begin{figure}[!b]
\centering
\includegraphics[width=\cmsFigWidth]{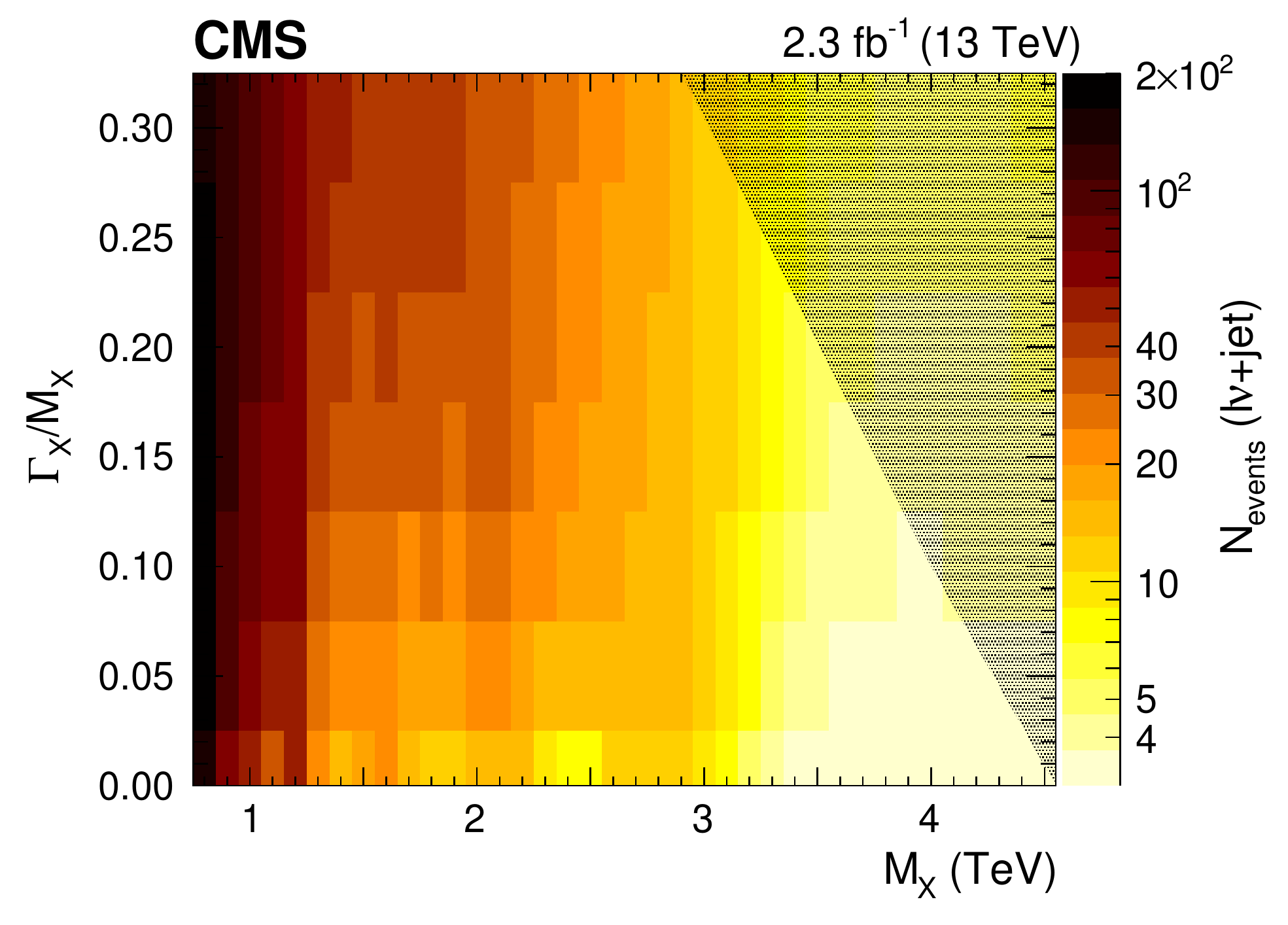}
\includegraphics[width=\cmsFigWidth]{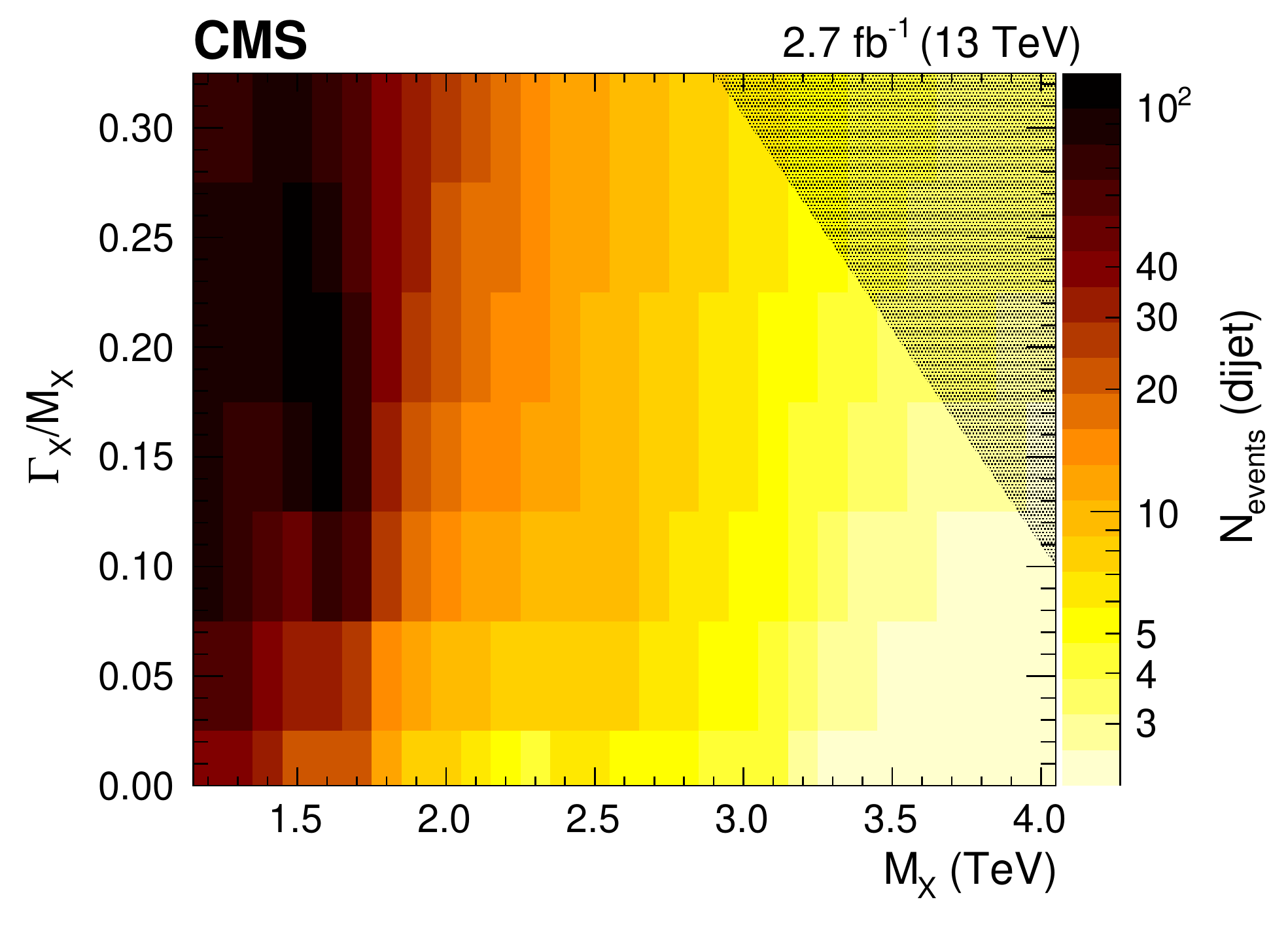}
\caption{Observed exclusion limits at 95\% CL on the number of events for
a $\PW\PV \to \lnujet$~(left) and a $\PV\PV \to \text{dijet}$~(right) resonance, as a function
of its mass and normalized width. The dark shaded area denotes the kinematic regime where the limit is valid only for the quark-antiquark annihilation processes.}
\label{fig:UL_scan2D}
\end{figure}

To validate the above procedure, the resulting parametrized efficiencies
(including the event veto efficiencies) are used to predict
the total efficiency for reconstructing
resonances of different spin and width.
The estimation is compared to the exact number obtained
from performing the baseline analysis directly on the simulated events.
In all cases, the agreement between the nominal and parametrized efficiencies are of the order of 10--20\% for the majority of the parameter space,
but grow up to 40\% for very low resonance masses, were migration effects over selection boundaries cannot be treated in our parametrization approach.
Various approximations and uncertainties contribute to the final additional systematic uncertainty in the efficiency;
the main ones are unaccounted correlations between the physical objects, statistical uncertainties due to
limited numbers of simulated events, and residual dependence on natural width.
We assign an additional systematic uncertainty which ranges from 20\% at high masses to 40\% at low masses
in the total signal efficiency for calculating the model-independent limits. This
additional systematic uncertainty addresses the remaining
imperfections in the parametrization of efficiencies.

Figure~\ref{fig:UL_scan2D}
shows the observed limits on the number of
events extracted from the simplified analysis, independently for the \lnujet{} and dijet
analyses, which are not combined in order to
avoid assumptions on branching fractions of a resonance decaying to both WW
and ZZ channels. The limits are calculated using an asymptotic
approximation of the $\mathrm{CL_S}$ method.
All systematic uncertainties considered in
the baseline analysis are included in the calculation of these
limits, together with the additional uncertainty related to the
approximations for parametrizing efficiencies.
The main features of the
observed limits presented above are still visible.
With increasing width, the overall sensitivity degrades.
The shaded area denotes where the limit is valid only for quark-antiquark annihilation processes,
because in this region the mass distribution resulting from
gluon-fusion processes can no longer be approximated by a peaking resonance.
\section{Summary}
\label{sec:summary}

A search has been presented for new resonances decaying to WW, ZZ, or WZ boson pairs
in which at least one of the bosons decays into quarks.
The final states involve dijet and \lnujet{} events with $\ell=\Pgm$ or \Pe.
The results include the $\PW\to\Pgt\Pgn$ contribution with subsequent decay $\Pgt\to\ell\Pgn\Pagn$.
The W and Z bosons that decay to quarks are selected by requiring a jet with mass compatible with the W or Z boson mass, respectively.
Additional information from jet substructure is used to suppress background from W+jets and multijet processes.
No evidence for a signal is found. In particular, the excesses at a resonance mass of 2\TeV observed in previous searches~\cite{Aad:2015owa,Khachatryan:2014hpa} are not confirmed. 
The result is interpreted as an upper limit on the
production cross section of a narrow-width resonance as a function its mass, in the
context of the bulk graviton model (with decays to WW or ZZ), heavy vector-triplet Models A and B, and \PWpr{} and \PZpr{} singlet models.
The upper limits are based on the statistical combination of the two channels.
For the heavy vector-triplet, we exclude \PWpr and \PZpr resonances with respective masses $<$2.0 and $<$1.6\TeV for Model A, $<$2.2 and $<$1.7\TeV for Model B.
Under the triplet hypothesis, spin-1 resonances with masses below 2.3 and 2.4\TeV are excluded for heavy vector-triplet Model A and B, respectively.
In the narrow-width bulk graviton model, cross sections are excluded in the range of 3--1200\unit{fb}.
This is the first search for a narrow-width bulk graviton with $\tilde{k} = 0.5$ at $\sqrt{s} = 13$\TeV.
Tabulated efficiencies for the reconstruction of the vector bosons within the kinematic acceptance of the analysis are also provided, allowing for a reintepretation of the exclusion limits in a generic phenomenological model.

\begin{acknowledgments}
We congratulate our colleagues in the CERN accelerator departments for the excellent performance of the LHC and thank the technical and administrative staffs at CERN and at other CMS institutes for their contributions to the success of the CMS effort. In addition, we gratefully acknowledge the computing centres and personnel of the Worldwide LHC Computing Grid for delivering so effectively the computing infrastructure essential to our analyses. Finally, we acknowledge the enduring support for the construction and operation of the LHC and the CMS detector provided by the following funding agencies: BMWFW and FWF (Austria); FNRS and FWO (Belgium); CNPq, CAPES, FAPERJ, and FAPESP (Brazil); MES (Bulgaria); CERN; CAS, MoST, and NSFC (China); COLCIENCIAS (Colombia); MSES and CSF (Croatia); RPF (Cyprus); SENESCYT (Ecuador); MoER, ERC IUT, and ERDF (Estonia); Academy of Finland, MEC, and HIP (Finland); CEA and CNRS/IN2P3 (France); BMBF, DFG, and HGF (Germany); GSRT (Greece); OTKA and NIH (Hungary); DAE and DST (India); IPM (Iran); SFI (Ireland); INFN (Italy); MSIP and NRF (Republic of Korea); LAS (Lithuania); MOE and UM (Malaysia); BUAP, CINVESTAV, CONACYT, LNS, SEP, and UASLP-FAI (Mexico); MBIE (New Zealand); PAEC (Pakistan); MSHE and NSC (Poland); FCT (Portugal); JINR (Dubna); MON, RosAtom, RAS, RFBR and RAEP (Russia); MESTD (Serbia); SEIDI and CPAN (Spain); Swiss Funding Agencies (Switzerland); MST (Taipei); ThEPCenter, IPST, STAR, and NSTDA (Thailand); TUBITAK and TAEK (Turkey); NASU and SFFR (Ukraine); STFC (United Kingdom); DOE and NSF (USA).

\hyphenation{Rachada-pisek} Individuals have received support from the Marie-Curie programme and the European Research Council and EPLANET (European Union); the Leventis Foundation; the A. P. Sloan Foundation; the Alexander von Humboldt Foundation; the Belgian Federal Science Policy Office; the Fonds pour la Formation \`a la Recherche dans l'Industrie et dans l'Agriculture (FRIA-Belgium); the Agentschap voor Innovatie door Wetenschap en Technologie (IWT-Belgium); the Ministry of Education, Youth and Sports (MEYS) of the Czech Republic; the Council of Science and Industrial Research, India; the HOMING PLUS programme of the Foundation for Polish Science, cofinanced from European Union, Regional Development Fund, the Mobility Plus programme of the Ministry of Science and Higher Education, the National Science Center (Poland), contracts Harmonia 2014/14/M/ST2/00428, Opus 2014/13/B/ST2/02543, 2014/15/B/ST2/03998, and 2015/19/B/ST2/02861, Sonata-bis 2012/07/E/ST2/01406; the Thalis and Aristeia programmes cofinanced by EU-ESF and the Greek NSRF; the National Priorities Research Program by Qatar National Research Fund; the Programa Clar\'in-COFUND del Principado de Asturias; the Rachadapisek Sompot Fund for Postdoctoral Fellowship, Chulalongkorn University and the Chulalongkorn Academic into Its 2nd Century Project Advancement Project (Thailand); and the Welch Foundation, contract C-1845. \end{acknowledgments}

\appendix
\section{Instructions and additional material for generic interpretation of the results}
\label{sec:mod-indep-instr}
This appendix presents a technical description of the procedure for calculating
the signal yield expected to be observed in the CMS detector in a scenario with
a new resonance, X, decaying to two vector bosons in the \lnujet{}
final state (WW, WZ), as well as the dijet final state (WW, WZ, and ZZ).
The efficiencies are calculated using the reference bulk graviton samples described in
Section~\ref{sec:simulatedsamples} and listed in Tables~\ref{tab:WWmuEff}--\ref{tab:WWallHadEff}.

These efficiencies can be applied to a generic model with the following procedure:
\begin{enumerate}
\item Generate a sample of events for a given mass and width of the X resonance;
the simulated process must include the decay of the X resonance
to leptons and quarks (including W$\to \tau\nu \to \ell\nu\nu\nu$ decays).
\item Split the sample into \lnujet{} and dijet decays.
\item Filter the events according to the criteria listed in
Table~\ref{tab:GeneratorSelectionsSemileptonic} (for \lnujet{} WW decays) and
Table~\ref{tab:AllHadGeneratorSelections} (for dijet WW decays).
If the resonance decays to $\PW\cPZ\to \ell\nu \Pq\Paq$,
the criteria for a hadronically decaying W boson in Table~\ref{tab:GeneratorSelectionsSemileptonic} should be applied to the generated hadronically decaying Z boson.
If the resonance decays to $\cPZ\PW$ or $\Zo\Zo$ in the dijet channel,
the criteria in Table~\ref{tab:AllHadGeneratorSelections} should be applied to the generated hadronically decaying Z bosons as well.
\item For each of the remaining events, calculate the efficiency
for reconstructing the channels $\Wo\to \mu\nu$ and $\Wo\to \tau\nu \to \mu \nu\nu \nu$,
and $\Wo\to \Pe \nu$ and $\Wo\to \tau \nu \to \Pe \nu\nu\nu$,
using Table~\ref{tab:WWmuEff}. The table provides the
efficiency parametrized as a function of \PT and $\eta$ of the W.
\item In a similar way, in the \lnujet{} channel calculate the efficiency of the hadronically decaying W or Z bosons using the values in Table~\ref{tab:WWhadEff}.
For the dijet decays, compute the efficiency for each boson from the values in Table~\ref{tab:WWallHadEff}.
\item Weight each accepted event with the product of the two efficiencies found
at steps 3 and 4. In the case of a X resonance decaying to WV (\lnujet channel),
also multiply by the combined efficiency of the second-lepton
and b jet vetoes, equal to $95\%$.
A correction factor amounting to 98\% should be applied to events in the dijet category to take into account the efficiency of the $\Delta\eta$ requirement.
\item The resulting sum of weighted events for the \lnujet{} and dijet subsamples, divided by the total number of
events, provides an approximation to the total efficiency for the given model in each of the two channels.
\end{enumerate}

\begin{table}[!htb]
\topcaption{Generator-level requirements for the \lnujet{} analysis, to be used for the computation of the efficiency parametrization.
The vector sum of the transverse neutrino momenta $\sum{\vec{p}_{\mathrm{T},\nu}}$ is taken over all the neutrinos in the final state, coming either from $\PW\to \ell\nu$ or $W\to \tau\nu \to \ell\nu\nu\nu$ decays with $\ell=\mu$ or \Pe.}
\label{tab:GeneratorSelectionsSemileptonic}
\centering
\begin{tabular}{lr}
\hline
Objects & Requirements \\
\hline
\hline
Muons & $\pt > 53\GeV$\\
            & $\abs{\eta}<$ 2.1\\
\hline
Electrons & $\pt > 120\GeV$\\
                &  $\abs{\eta}<$ 2.5\\
\hline
$\sum{\vec{p}_{\mathrm{T},\nu}}$ & $\pt>40\GeV$ (muon channel)\\
                             & $\pt>80\GeV$ (electron channel)\\
\hline
W $\to\ell\nu$ or W $\to\tau\nu\to\ell\nu\nu\nu$ & $\pt >200\GeV$\\
\hline
V $\to\qqbar$ & $\pt >200\GeV$ \\
  			     & $\abs{\eta}<2.4$ \\
\hline
WV system & $0.7 < m_\mathrm{WV} < 5.0\TeV$\\
          & $\Delta \phi(\Vo_{\Pq\Paq},\Wo_{l\nu})>2$\\
          & $\Delta \phi(\Vo_{\Pq\Paq},\sum{\vec{p}_{\mathrm{T},\nu}})>2$\\
          & $\Delta R(\Vo_{\Pq\Paq},\ell)> \pi/2$\\
\hline
\end{tabular}
\end{table}

\begin{table}[!htb]
\centering
\topcaption{Generator-level requirements for the dijet analysis, to be used for the computation of the efficiency parametrization.}
\label{tab:AllHadGeneratorSelections}
\begin{tabular}{lr}
\hline
Objects & Requirements \\
\hline
\hline
V $\to\qqbar$ & $\pt >200\GeV$\\
			     & $\abs{\eta}<2.4$ \\
\hline
VV system & $ m_\mathrm{VV}> 1\TeV$\\
          & $\abs{\eta_\mathrm{V_1}-\eta_\mathrm{V_2} }<1.3$\\
\hline
\end{tabular}
\end{table}

The final numbers of events can be directly compared to the observed
limits in Fig.~\ref{fig:UL_scan2D} and Table~\ref{tab:simplim_WW}, in order to assess the exclusion power of the present data with respect to the model considered.

The numbers provided refer to longitudinally polarized bosons. For
transversely polarized bosons that decay leptonically, the same
numbers are valid, as long as they are applied after the kinematic
acceptance requirements. If the boson decays to quarks and has a transverse
polarization, the efficiency must be scaled down by a factor of 0.76 for each hadronically decaying boson in the event.

\begin{table}[!htb]
\centering
\topcaption{Reconstruction and identification efficiency for the (upper table) $\Wo\to \mu \nu$ and $\Wo \to \tau \nu \to \mu \nu\nu\nu$, and (lower table) $\Wo\to \Pe \nu$ and $\Wo \to \tau \nu \to \Pe \nu\nu\nu$ decays
as function of generated $\pt^\Wo$ and $\abs{\eta_\Wo}$. Uncertainties in the efficiencies are included in the generic limit calculation as discussed in the text.}
\label{tab:WWmuEff}
\centering
\resizebox{\textwidth}{!}{
\begin{tabular}{l c c c c c c c c c}
\hline
\multicolumn{10}{c}{$\Wo\to \mu \nu$ and $\Wo \to \tau \nu \to \mu \nu\nu\nu$}\\[1pt]
${\pt^\Wo}$ {range} (\GeVns{}) & \multicolumn{9}{c}{${\abs{\eta_\Wo}}$ {range}}\\[5pt]
 & 0--0.2 & 0.2--0.4 & 0.4--0.6 & 0.6--0.8 & 0.8--1.0 & 1--1.25 & 1.25--1.5 & 1.5--2.0 & 2--2.5 \\
\hline
\hline
200--250 & 0.82 & 0.79 & 0.79 & 0.80 & 0.85 & 0.82 & 0.81 & 0.82 & 0.78 \\
250--300 & 0.89 & 0.90 & 0.88 & 0.86 & 0.90 & 0.86 & 0.91 & 0.86 & 0.91 \\
300--400 & 0.90 & 0.89 & 0.90 & 0.90 & 0.89 & 0.90 & 0.89 & 0.90 & 0.87 \\
400--500 & 0.88 & 0.89 & 0.91 & 0.90 & 0.88 & 0.89 & 0.90 & 0.89 & 0.89 \\
500--600 & 0.90 & 0.90 & 0.92 & 0.90 & 0.88 & 0.89 & 0.91 & 0.87 & 0.88 \\
600--700 & 0.91 & 0.90 & 0.92 & 0.91 & 0.88 & 0.90 & 0.92 & 0.88 & 0.87 \\
700--800 & 0.91 & 0.89 & 0.92 & 0.91 & 0.89 & 0.89 & 0.91 & 0.90 & 0.82 \\
800--1000 & 0.92 & 0.89 & 0.92 & 0.91 & 0.88 & 0.88 & 0.90 & 0.88 & 0.94 \\
1000--1200 & 0.91 & 0.89 & 0.92 & 0.91 & 0.89 & 0.88 & 0.89 & 0.85 & 0.75 \\
1200--1500 & 0.91 & 0.88 & 0.92 & 0.91 & 0.87 & 0.87 & 0.89 & 0.87 &\NA  \\
1500--2000 & 0.90 & 0.87 & 0.92 & 0.91 & 0.86 & 0.88 & 0.87 &\NA  &\NA  \\
2000--2500 & 0.91 & 0.86 & 0.91 & 0.90 & 0.83 & 0.82 &\NA  &\NA  &\NA  \\
2500--3000 & 0.88 & 0.79 & 0.90 & 0.82 &\NA  &\NA  &\NA  &\NA  &\NA  \\
3000--4000 & 0.78 & 0.88 & 0.80 & 1.00 &\NA  &\NA  &\NA  &\NA  &\NA  \\
\hline
\multicolumn{10}{c}{}\\[0.1pt]
\multicolumn{10}{c}{$\Wo\to \Pe \nu$ and $\Wo \to \tau \nu \to \Pe \nu\nu\nu$}\\
\multicolumn{10}{c}{}\\[0.1pt]
\hline
\hline
200--250 & 0.78 & 0.75 & 0.82 & 0.81 & 0.79 & 0.80 & 0.71 & 0.79 & 0.68 \\
250--300 & 0.79 & 0.79 & 0.77 & 0.80 & 0.78 & 0.82 & 0.79 & 0.73 & 0.78 \\
300--400 & 0.82 & 0.82 & 0.82 & 0.83 & 0.82 & 0.82 & 0.80 & 0.81 & 0.80 \\
400--500 & 0.82 & 0.82 & 0.81 & 0.84 & 0.81 & 0.81 & 0.82 & 0.82 & 0.80 \\
500--600 & 0.83 & 0.83 & 0.84 & 0.84 & 0.83 & 0.81 & 0.82 & 0.84 & 0.85 \\
600--700 & 0.83 & 0.84 & 0.84 & 0.83 & 0.85 & 0.84 & 0.82 & 0.84 & 0.88 \\
700--800 & 0.84 & 0.83 & 0.84 & 0.85 & 0.84 & 0.84 & 0.82 & 0.82 & 0.94 \\
800--1000 & 0.83 & 0.84 & 0.84 & 0.84 & 0.85 & 0.86 & 0.82 & 0.85 & 0.78 \\
1000--1200 & 0.83 & 0.84 & 0.84 & 0.83 & 0.84 & 0.85 & 0.84 & 0.86 & 0.33 \\
1200--1500 & 0.84 & 0.84 & 0.84 & 0.84 & 0.85 & 0.84 & 0.85 & 0.81 &\NA  \\
1500--2000 & 0.83 & 0.85 & 0.84 & 0.84 & 0.86 & 0.84 & 0.86 & 0.95 &\NA  \\
2000--2500 & 0.83 & 0.85 & 0.84 & 0.85 & 0.84 & 0.79 &\NA  &\NA  &\NA  \\
2500--3000 & 0.78 & 0.82 & 0.78 & 0.69 &\NA  &\NA  &\NA  &\NA  &\NA  \\
3000--4000 & 0.80 & 0.81 & 0.67 & 1.00 &\NA  &\NA  &\NA  &\NA  &\NA  \\
\hline
\end{tabular}
}
\end{table}

\begin{table}[!htb]
\centering
\topcaption{Reconstruction and identification efficiency for the (upper table) $\Wo_{\text{L}}\to\qqbar$ and (lower table) $\Zo_{\text{L}}\to\qqbar$ decay as a function of generated $\pt^\Vo$ and $\abs{\eta_\Vo}$ applying the V tagging requirements used in the \lnujet{} analysis ($\nsubj<0.6$). Uncertainties in the efficiencies are included in the generic limit calculation as discussed in the text.}
\label{tab:WWhadEff}
\resizebox{\textwidth}{!}{
\begin{tabular}{l c c c c c c c c c}
\hline
\multicolumn{10}{c}{$\Wo_{\text{L}}\to\qqbar$}\\[1pt]
${\pt^\Wo}$ {range} (\GeVns{}) & \multicolumn{9}{c}{${\abs{\eta_\Wo}}$ {range}}\\[5pt]
 & 0--0.2 & 0.2--0.4 & 0.4--0.6 & 0.6--0.8 & 0.8--1.0 & 1.0--1.25 & 1.25--1.5 & 1.5--2.0 & 2.0--2.5 \\
\hline
\hline
200--250 & 0.31 & 0.36 & 0.33 & 0.28 & 0.37 & 0.38 & 0.30 & 0.25 & 0.26 \\
250--300 & 0.54 & 0.48 & 0.57 & 0.46 & 0.50 & 0.54 & 0.47 & 0.48 & 0.56 \\
300--400 & 0.71 & 0.70 & 0.72 & 0.70 & 0.70 & 0.65 & 0.66 & 0.63 & 0.59 \\
400--500 & 0.65 & 0.65 & 0.66 & 0.64 & 0.64 & 0.67 & 0.62 & 0.63 & 0.70 \\
500--600 & 0.72 & 0.71 & 0.73 & 0.72 & 0.73 & 0.70 & 0.66 & 0.69 & 0.72 \\
600--700 & 0.74 & 0.75 & 0.74 & 0.73 & 0.72 & 0.71 & 0.71 & 0.72 & 0.78 \\
700--800 & 0.73 & 0.74 & 0.73 & 0.75 & 0.72 & 0.71 & 0.67 & 0.68 & 0.65 \\
800--1000 & 0.73 & 0.74 & 0.74 & 0.74 & 0.73 & 0.71 & 0.65 & 0.66 & 0.62 \\
1000--1200 & 0.69 & 0.71 & 0.71 & 0.71 & 0.69 & 0.66 & 0.57 & 0.65 & 0.67 \\
1200--1500 & 0.68 & 0.69 & 0.69 & 0.70 & 0.68 & 0.67 & 0.54 & 0.63 & \NA  \\
1500--2000 & 0.69 & 0.69 & 0.69 & 0.68 & 0.67 & 0.65 & 0.47 & 0.11 & \NA  \\
2000--2500 & 0.68 & 0.68 & 0.69 & 0.69 & 0.67 & 0.66 & 0.50 & \NA  & \NA  \\
2500--3000 & 0.76 & 0.63 & 0.77 & 0.53 & 0.67 & \NA  & \NA  & \NA  & \NA  \\
3000--4000 & 0.77 & 0.43 & 1.00 & 1.00 & \NA  & \NA  & \NA  & \NA  & \NA  \\
\hline
\multicolumn{10}{c}{}\\[0.1pt]
\multicolumn{10}{c}{$\Zo_{\text{L}}\to\qqbar$}\\
\multicolumn{10}{c}{}\\[0.1pt]
\hline
\hline
200--250 & 0.26 & 0.48 & 0.27 & 0.37 & 0.33 & 0.41 & 0.37 & 0.36 & 0.28 \\
250--300 & 0.64 & 0.56 & 0.62 & 0.58 & 0.60 & 0.57 & 0.56 & 0.61 & 0.53 \\
300--400 & 0.76 & 0.75 & 0.77 & 0.75 & 0.74 & 0.72 & 0.71 & 0.73 & 0.67 \\
400--500 & 0.75 & 0.75 & 0.76 & 0.74 & 0.76 & 0.77 & 0.73 & 0.74 & 0.71 \\
500--600 & 0.80 & 0.81 & 0.82 & 0.80 & 0.78 & 0.79 & 0.76 & 0.78 & 0.73 \\
600--700 & 0.81 & 0.83 & 0.80 & 0.81 & 0.82 & 0.80 & 0.76 & 0.77 & 0.72 \\
700--800 & 0.81 & 0.80 & 0.79 & 0.79 & 0.80 & 0.78 & 0.74 & 0.75 & 0.77 \\
800--1000 & 0.81 & 0.81 & 0.81 & 0.81 & 0.79 & 0.77 & 0.72 & 0.74 & 0.72 \\
1000--1200 & 0.78 & 0.78 & 0.79 & 0.78 & 0.77 & 0.75 & 0.66 & 0.71 & 0.77 \\
1200--1500 & 0.77 & 0.77 & 0.76 & 0.77 & 0.75 & 0.73 & 0.60 & 0.65 & 0.66 \\
1500--2000 & 0.74 & 0.74 & 0.73 & 0.74 & 0.72 & 0.67 & 0.52 & 0.57 &\NA  \\
2000--2500 & 0.72 & 0.73 & 0.73 & 0.73 & 0.69 & 0.66 & 0.53 &\NA  &\NA  \\
2500--3000 & 0.80 & 0.85 & 0.69 & 0.76 &\NA  &\NA  &\NA  &\NA  &\NA  \\
3000--4000 & 1.0 & 0.50 & 1.0 &\NA  &\NA  &\NA  &\NA  &\NA  &\NA  \\
\hline
\end{tabular}
}
\end{table}

\begin{table}[!htb]
\centering
\caption{Reconstruction and identification efficiency for the (upper table) $\Wo_{\text{L}}\to\qqbar$ and (lower table) $\Zo_{\text{L}}\to\qqbar$ decays as a function of generated $\pt^\Vo$ and $\abs{\eta_\Vo}$ applying the V tagging requirements used in the dijet analysis ($\nsubj<0.45$). Uncertainties in the efficiencies are included in the generic limit calculation as discussed in the text.}
\label{tab:WWallHadEff}
\centering
\resizebox{\textwidth}{!}{
\begin{tabular}{l c c c c c c c c c c }
\hline
\multicolumn{11}{c}{$\Wo_{\text{L}}\to\qqbar$}\\[1pt]
${\pt^\Wo}$ {range} (\GeVns{}) & \multicolumn{10}{c}{${\abs{\eta_\Wo}}$ {range}}\\[5pt]
 & 0.0--0.2 & 0.2--0.3 & 0.3--0.4 & 0.4--0.6 & 0.6--0.8 & 0.8--1.0 & 1.00--1.25 & 1.2--1.5 & 1.5--2.0 & 2.0--2.4 \\
\hline
\hline
200--250 & 0.27 & 0.34 & 0.23 & 0.25 & 0.35 & 0.32 & 0.31 & 0.30 & 0.32 \\
250--300 & 0.55 & 0.50 & 0.55 & 0.51 & 0.54 & 0.58 & 0.52 & 0.56 & 0.54 \\
300--400 & 0.74 & 0.74 & 0.75 & 0.73 & 0.73 & 0.69 & 0.69 & 0.67 & 0.63 \\
400--500 & 0.69 & 0.68 & 0.70 & 0.69 & 0.68 & 0.69 & 0.65 & 0.65 & 0.71 \\
500--600 & 0.72 & 0.72 & 0.74 & 0.73 & 0.74 & 0.70 & 0.66 & 0.70 & 0.75 \\
600--700 & 0.74 & 0.75 & 0.75 & 0.74 & 0.73 & 0.72 & 0.71 & 0.73 & 0.78 \\
700--800 & 0.74 & 0.75 & 0.74 & 0.75 & 0.73 & 0.72 & 0.68 & 0.69 & 0.66 \\
800--1000 & 0.74 & 0.75 & 0.75 & 0.75 & 0.73 & 0.71 & 0.66 & 0.66 & 0.58 \\
1000--1200 & 0.70 & 0.71 & 0.72 & 0.72 & 0.69 & 0.67 & 0.59 & 0.63 & 0.40 \\
1200--1500 & 0.69 & 0.70 & 0.70 & 0.70 & 0.68 & 0.65 & 0.54 & 0.59 &\NA  \\
1500--2000 & 0.68 & 0.69 & 0.68 & 0.68 & 0.67 & 0.65 & 0.47 &\NA  &\NA  \\
2000--2500 & 0.69 & 0.69 & 0.69 & 0.69 & 0.67 & 0.69 &\NA  &\NA  &\NA  \\
2500--3000 & 0.74 & 0.66 & 0.73 & 0.60 &\NA  &\NA  &\NA  &\NA  &\NA  \\
3000--4000 & 0.74 & 0.74 & 1.00 & 1.00 &\NA  &\NA  &\NA  &\NA  &\NA  \\
\hline
\multicolumn{11}{c}{}\\[0.1pt]
\multicolumn{11}{c}{$\Zo_{\text{L}}\to\qqbar$}\\
\multicolumn{10}{c}{}\\[0.1pt]
\hline
\hline
200--250 & \NA  & \NA  & 0.25 & \NA  & \NA  & 0.50 & \NA  & 0.50 & \NA  & \NA  \\
250--300 & 0.30 & \NA  & 0.33 & 0.25 & 0.18 & \NA  & 0.33 & 0.67 & \NA  & \NA  \\
300--350 & 0.46 & 0.15 & 0.33 & 0.44 & 0.45 & 0.21 & 0.41 & 0.12 & 0.44 & \NA  \\
350--400 & 0.38 & 0.40 & 0.47 & 0.47 & 0.43 & 0.41 & 0.26 & 0.35 & 0.36 & 0.50 \\
400--500 & 0.50 & 0.46 & 0.51 & 0.47 & 0.45 & 0.51 & 0.41 & 0.38 & 0.45 & 0.59 \\
500--600 & 0.59 & 0.60 & 0.61 & 0.60 & 0.58 & 0.55 & 0.52 & 0.44 & 0.51 & 0.67 \\
600--700 & 0.63 & 0.61 & 0.62 & 0.59 & 0.59 & 0.56 & 0.50 & 0.45 & 0.48 & 0.53 \\
700--800 & 0.60 & 0.62 & 0.61 & 0.60 & 0.60 & 0.58 & 0.50 & 0.40 & 0.41 & 0.75 \\
800--1000 & 0.60 & 0.60 & 0.59 & 0.60 & 0.56 & 0.52 & 0.48 & 0.38 & 0.46 & 1.00 \\
1000--1200 & 0.55 & 0.52 & 0.57 & 0.52 & 0.53 & 0.48 & 0.43 & 0.25 & 0.40 & 1.00 \\
1200--1500 & 0.53 & 0.52 & 0.53 & 0.52 & 0.50 & 0.44 & 0.39 & 0.25 & 0.16 & \NA  \\
1500--2000 & 0.49 & 0.49 & 0.48 & 0.47 & 0.46 & 0.42 & 0.34 & 0.24 & \NA  & \NA  \\
2000--2500 & 0.47 & 0.49 & 0.48 & 0.44 & 0.43 & 0.42 & 0.41 & 0.33 & \NA  & \NA  \\
2500--3000 & 0.43 & 0.36 & 0.47 & 0.47 & 0.38 & 0.17 & \NA  & \NA  & \NA  & \NA  \\
3000--4000 & 0.44 & 0.50 & \NA  & \NA  & \NA  & \NA  & \NA  & \NA  & \NA  & \NA  \\
\hline
\end{tabular}
}
\end{table}

\begin{table}[!htb]
\centering
\topcaption{Simplified limits on the number of visible events from generic resonances decaying to pairs of V bosons in the \lnujet (left) and dijet (right) channels as a function of resonance mass, $M_\mathrm{X}$, and normalized width, $\Gamma_\mathrm{X}/M_\mathrm{X}$. Shown are limits on the visible number of events at 95\% CL using the asymptotic $\mathrm{CL_S}$ approach. Results with $\Gamma_\mathrm{X}/M_\mathrm{X}=0$ are obtained using the resolution function only.}
\label{tab:simplim_WW}
\resizebox{\textwidth}{!}{
\begin{tabular}{c c c c c c c c | c c c c c c c }
\multirow{3}{*}{$M_\mathrm{X}$ (\TeVns{}) }  & \multicolumn{7}{c|}{\lnujet channel} & \multicolumn{7}{c}{dijet channel}\\[5pt]
   &\multicolumn{7}{c|}{${\Gamma_\mathrm{X} / M_\mathrm{X}}$} &\multicolumn{7}{c}{${\Gamma_\mathrm{X} / M_\mathrm{X}}$}\\[5pt]
   & 0.00 & 0.05 & 0.10 & 0.15 & 0.20 & 0.25 & 0.30 & 0.00 & 0.05 & 0.10 & 0.15 & 0.20 & 0.25 & 0.30 \\
\hline
\hline
0.8 & 139.9  & 173.5  & 189.2  & 192.7  & 185.7  & 173.1  & 157.8 & \NA  & \NA  & \NA  & \NA  & \NA  & \NA  & \NA\\
0.9 & 66.9  & 87.8  & 104.4  & 115.5  & 120.4 & 120.6  & 117.1 & \NA  & \NA  & \NA  & \NA  & \NA  & \NA  & \NA\\
1.0 & 46.9  & 61.4  & 72.4  & 81.6  & 87.9 & 91.0  & 91.4 & \NA  & \NA  & \NA  & \NA  & \NA  & \NA  & \NA\\
1.1 & 35.2  & 47.1  & 58.2  & 66.7  & 72.2 & 75.3  & 76.3 & \NA  & \NA  & \NA  & \NA  & \NA  & \NA  & \NA\\
1.2 & 50.7  & 56.5  & 59.7  & 62.0  & 63.8 & 65.1  & 65.9 & 38.3  & 61.7  & 88.6  & 84.8  & 84.3 & 82.8  & 78.1\\
1.3 & 22.7  & 29.4  & 34.9  & 40.4  & 45.5 & 49.9  & 53.0 & 39.6  & 54.9  & 68.9  & 77.8  & 82.2 & 83.0  & 79.0\\
1.4 & 15.1  & 20.5  & 26.3  & 32.1  & 37.9 & 43.0  & 46.9 & 29.8  & 41.9  & 57.3  & 66.7  & 82.7 & 86.3  & 85.7\\
1.5 & 18.2  & 22.4  & 27.1  & 32.1  & 37.2 & 41.7  & 45.1 & 19.7  & 31.0  & 45.6  & 89.1  & 127.4 & 116.0  & 93.4\\
1.6 & 20.1  & 24.1  & 28.4  & 33.4  & 38.3 & 42.3  & 44.9 & 22.4  & 34.0  & 65.7  & 114.8  & 100.5 & 90.1  & 77.3\\
1.7 & 14.2  & 19.0  & 24.4  & 30.6  & 36.7 & 41.3  & 44.0 & 22.1  & 29.1  & 57.6  & 70.9  & 70.9 & 64.6  & 57.2\\
1.8 & 11.8  & 17.7  & 24.5  & 31.6  & 37.0 & 40.0  & 40.6 & 13.0  & 15.4  & 24.2  & 34.6  & 40.4 & 41.1  & 39.7\\
1.9 & 11.6  & 16.6  & 23.1  & 29.8  & 35.1 & 38.4  & 39.7 & 7.7  & 11.8  & 17.2  & 23.7  & 27.8 & 29.3  & 29.5\\
2.0 & 14.7  & 20.4  & 26.7  & 32.0  & 35.2 & 36.8  & 37.2 & 7.7  & 10.6  & 14.5  & 18.7  & 21.5 & 23.1  & 23.9\\
2.1 & 15.4  & 20.8  & 26.4  & 30.6  & 32.7 & 33.6  & 33.9 & 6.2  & 9.0  & 12.5  & 15.6  & 17.5 & 18.8  & 19.5\\
2.2 & 13.2  & 18.5  & 23.9  & 27.5  & 29.4 & 30.2  & 30.6 & 5.1  & 7.8  & 10.9  & 13.4  & 15.1 & 15.9  & 16.5\\
2.3 & 9.8  & 15.4  & 20.7  & 24.2  & 26.1 & 27.1  & 27.3 & 4.6  & 7.8  & 10.5  & 12.2  & 13.2 & 13.8  & 14.3\\
2.4 & 7.9  & 13.3  & 18.4  & 21.4  & 23.0 & 24.2  & 25.1 & 5.9  & 8.4  & 10.2  & 11.1  & 11.8 & 12.2  & 12.6\\
2.5 & 8.5  & 13.7  & 17.4  & 19.5  & 20.6 & 21.6  & 22.6 & 6.4  & 8.4  & 9.5  & 10.1  & 10.6  & 11.0  & 11.3\\
2.6 & 11.0  & 14.6  & 16.7  & 18.0  & 18.8 & 19.5  & 20.3 & 5.5  & 7.8  & 8.7  & 9.2  & 9.6  & 9.9  & 10.2\\
2.7 & 11.9  & 14.6  & 16.0  & 16.8  & 17.3 & 17.7  & 18.4 & 4.8  & 7.0  & 7.8  & 8.3  & 8.6  & 8.9  & 9.2\\
2.8 & 12.3  & 14.1  & 15.0  & 15.5  & 16.0 & 16.2  & 16.7 & 4.8  & 6.2  & 6.8  & 7.3  & 7.7  & 8.0  & 8.3\\
2.9 & 11.9  & 13.1  & 13.8  & 14.3  & 14.6 & 14.9  & 15.2 & 4.6  & 5.5  & 6.1  & 6.5  & 6.9  & 7.2  & 7.4\\
3.0 & 9.5  & 11.0  & 11.7  & 12.0  & 12.4 & 12.5  & 12.6 & 4.5  & 5.1  & 5.6  & 5.9  & 6.3  & 6.6  & 6.9\\
3.1 & 7.5  & 9.2  & 10.1  & 10.7  & 11.2 & 11.6  & 11.9 & 4.1  & 4.5  & 4.9  & 5.3  & 5.6  & 5.9  & 6.2\\
3.2 & 5.6  & 7.1  & 8.0  & 8.8  & 9.4  & 9.9  & 10.3 & 2.9  & 3.7  & 4.2  & 4.6  & 5.0  & 5.3  & 5.6\\
3.3 & 4.0  & 5.3  & 6.2  & 7.0  & 7.7  & 8.3  & 8.8 & 2.4  & 3.1  & 3.6  & 4.0  & 4.4  & 4.8  & 5.1\\
3.4 & 3.4  & 4.3  & 5.1  & 5.8  & 6.5  & 7.1  & 7.6 & 2.3  & 2.7  & 3.1  & 3.5  & 4.0  & 4.3  & 4.6\\
3.5 & 3.2  & 3.9  & 4.5  & 5.1  & 5.6  & 6.2  & 6.8 & 2.2  & 2.5  & 2.8  & 3.2  & 3.6  & 3.9  & 4.2\\
3.6 & 3.0  & 3.6  & 4.1  & 4.6  & 5.1  & 5.6  & 6.1 & 2.2  & 2.5  & 2.7  & 3.0  & 3.5  & 3.7  & 4.0\\
3.7 & 3.0  & 3.5  & 3.9  & 4.3  & 4.7  & 5.2  & 5.7 & 2.2  & 2.4  & 2.6  & 2.9  & 3.3  & 3.5  & 3.8\\
4.0 & 3.1  & 3.4  & 3.7  & 4.0  & 4.4  & 4.8  & 5.3 & 2.1  & 2.2  & 2.3  & 2.6  & 3.0  & 3.2  & 3.4\\
4.1 & 3.3  & 3.6  & 3.9  & 4.2  & 4.6  & 5.0  & 5.5 & \NA  & \NA  & \NA  & \NA  & \NA  & \NA  & \NA\\
4.5 & 3.4  & 3.7  & 4.0  & 4.3  & 4.8  & 5.3  & 6.1 & \NA  & \NA  & \NA  & \NA  & \NA  & \NA  & \NA\\
\hline
\end{tabular}}
\end{table}

\clearpage
\bibliography{auto_generated}

\cleardoublepage \section{The CMS Collaboration \label{app:collab}}\begin{sloppypar}\hyphenpenalty=5000\widowpenalty=500\clubpenalty=5000\input{B2G-16-004-authorlist.tex}\end{sloppypar}
\end{document}

%% file: B2G-16-004-authorlist.tex
\textbf{Yerevan Physics Institute,  Yerevan,  Armenia}\\*[0pt]
A.M.~Sirunyan, A.~Tumasyan
\vskip\cmsinstskip
\textbf{Institut f\"{u}r Hochenergiephysik,  Wien,  Austria}\\*[0pt]
W.~Adam, E.~Asilar, T.~Bergauer, J.~Brandstetter, E.~Brondolin, M.~Dragicevic, J.~Er\"{o}, M.~Flechl, M.~Friedl, R.~Fr\"{u}hwirth\cmsAuthorMark{1}, V.M.~Ghete, C.~Hartl, N.~H\"{o}rmann, J.~Hrubec, M.~Jeitler\cmsAuthorMark{1}, A.~K\"{o}nig, I.~Kr\"{a}tschmer, D.~Liko, T.~Matsushita, I.~Mikulec, D.~Rabady, N.~Rad, B.~Rahbaran, H.~Rohringer, J.~Schieck\cmsAuthorMark{1}, J.~Strauss, W.~Waltenberger, C.-E.~Wulz\cmsAuthorMark{1}
% ...
\vskip\cmsinstskip
% ...
1:~~Also at Vienna University of Technology, Vienna, Austria\\
% ...
